\documentclass[fleqn,usenatbib]{mnras}

\usepackage{newtxtext,newtxmath}

\usepackage[T1]{fontenc}
\usepackage{longtable}
\usepackage[justification=centering]{caption}
\DeclareRobustCommand{\VAN}[3]{#2}
\let\VANthebibliography\thebibliography
\def\thebibliography{\DeclareRobustCommand{\VAN}[3]{##3}\VANthebibliography}

\usepackage{graphicx}	
\usepackage{amsmath}	

\title[Chemical models of cyanamide and carbodiimide]{Chemical models of adenine precursors cyanamide and carbodiimide in the interstellar medium}

\author[X. Zhang et al.]{
Xia Zhang,$^{1,2}$
Donghui Quan,$^{3,1}$\thanks{E-mail:donghui.quan@zhejianglab.com}
Runxia Li,$^{2}$
Jarken Esimbek,$^{1,2,4}$
Long-Fei Chen,$^{3}$
\newauthor
Guoming Zhao$^{5}$
and Yan Zhou$^{6}$
\\
$^{1}$ Xinjiang Astronomical Observatory, Chinese Academy of Sciences, 150 Science 1-Street, Urumqi, Xinjiang 830011, China\\
$^{2}$Xinjiang Key Laboratory of Radio Astrophysics, 150 Science1-Street, Urumqi 830011, China\\
$^{3}$Research Center for Intelligent Computing Platforms, Zhejiang Laboratory, Hangzhou 311100, China\\
$^{4}$Key Laboratory of Radio Astronomy, Chinese Academy of Sciences, Urumqi 830011, China\\
$^{5}$School of Science, Jilin Institute of Chemical Technology, Jilin, Jilin 132022, China\\
$^{6}$BinZhou University, Huanghe Road, Binzhou City, Shandong, 256600, China}

\date{Accepted XXX. Received YYY; in original form ZZZ}

\pubyear{2022}

\begin{document}
\label{firstpage}
\pagerange{\pageref{firstpage}--\pageref{lastpage}}
\maketitle

\begin{abstract}
Cyanamide (NH$_2$CN) and its isomer, carbodiimide (HNCNH), may form adenine in the interstellar medium (ISM) via a series of reactions. Therefore, they are considered key prebiotic molecules in the study of the origin of life. We used the three-phase NAUTILUS chemical code, which includes the gas, the dust surface, and the icy mantle, to investigate the formation and destruction of cyanamide and carbodiimide. We added over 200 new chemical reactions of the two isomers and related species, and established a relatively complete network. We applied cold core, hot corino/core and shock models to simulate the different physical environments, and found that the two isomers are mainly produced by the free radical reactions on grain surfaces. Our simulated results suggest that cyanamide and carbodiimide molecules come from surface chemistry at early evolutionary stages. Then they are released back to the gas phase, either by thermal process (in hot cores, hot corinos) or shock-induced desorption (in shock regions). We speculate that it is an inefficient route to form a tautomer of adenine by starting from molecules cyanoacetylene (C$_3$NH), cyanamide and carbodiimide in ISM.
\end{abstract}

\begin{keywords}
astrochemistry -- molecular processes -- ISM: abundances -- ISM: molecules.
\end{keywords}



\section{Introduction}

How life on Earth began remains an unexplained scientific problem. Prebiotic chemistry addresses this issue through the aspects of theory, experiment and observation. However, large molecules of life such as ribonucleic acid RNA break down easily and are difficult to survive in the harsh interstellar environment. Therefore, they may be studied indirectly by their precursors: relatively small prebiotic molecules. Molecules containing cyano (CN) functional group are regarded as intermediates for the formation of purines and proteins \citep{becker2013}. Adenine is one of purine of DNA and RNA nucleobases and is always presumed to be the product of HCN pentamer reaction \citep{chakrabarti2000}. However, \citet{smith2001} found the reaction from HCN to adenine to be inefficient under interstellar medium (ISM) conditions. \citet{merz2014} provided a new route for adenine synthesis, using the concept of retrosynthetic analysis to create a tautomer of adenine, starting from molecules cyanoacetylene (C$_3$NH), carbodiimide(HNCNH) and its isomer cyanamide (NH$_2$CN). \citet{chakrabarti2015} proposed that observers may look in the ISM for the two precursors C$_3$NH and HNCNH which are equally important for predicting abundances of adenine. Both C$_3$NH and HNCNH have been observed in the ISM \citep{kawaguchi1992, mcguire2012}.

\begin{figure}
\centerline{\includegraphics[scale=0.5]{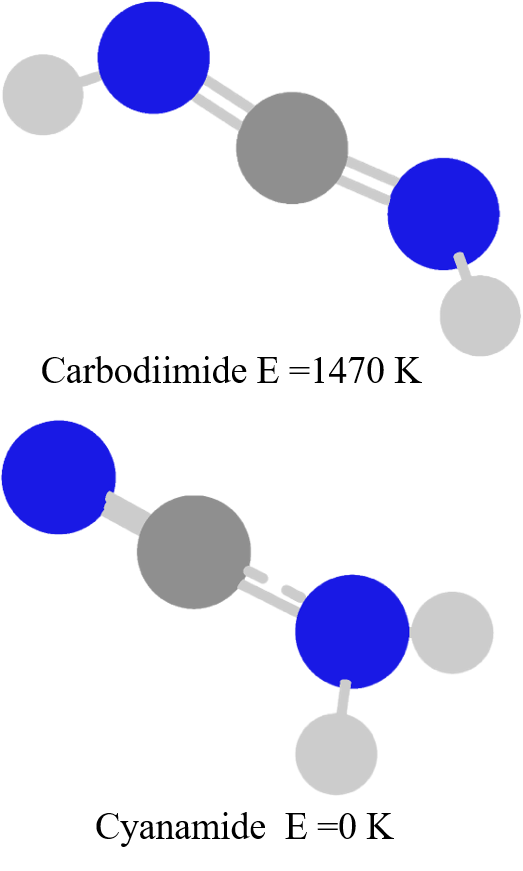}}
\caption{Optimized structures of the two isomers. Energies have been calculated at the CCSD(T)/aug-cc-pvtz//B2PLYPD3/may-cc-pvtz level. Blue, gray, white spheres correspond to nitrogen, carbon, and hydrogen atoms, respectively.}
\label{fig:molecular_structure}
\end{figure}

NH$_2$CN is one of the simplest organic molecules featuring cyano and amide groups. It is also the most stable of the isomers with the formula CH$_2$N$_2$. HNCNH is the second one, which is less stable than NH$_2$CN by about 4 kcal mol$^{-1}$ and therefore it is expected to be present in a smaller amount at thermal equilibrium \citep{jabs1999}. The structural images of two isomers are shown in Fig. \ref{fig:molecular_structure}. NH$_2$CN is recognized as an important prebiotic molecule in the study of the origin of life \citep{steinman1964, brack1999}. It can be converted into urea in liquid water \citep{kilpatrick1947}, while HNCNH can lead to isourea, a urea tautomer \citep{tordini2003}. Urea has been observed to be present in Sgr B2(N1) \citep{belloche2019}. However, \citet{duvernay2004, duvernay2005} did not detect the hydrolysis reaction of cyanamide or carbodiimide to form urea or isourea at low temperature ($\le$180 K) in their experimental study. Molecules with HNCNH moiety (-NCN-) are considered important condensing agents that are able to assemble amino acids into peptides in liquid water \citep{williams1981}. 

\begin{table*}
 \caption{Derived physical parameters of NH$_2$CN toward different types of sources.}
 \label{tab:molecular-column}
 \begin{tabular}{p{2.6cm}p{1.2cm}p{2.1cm}p{2cm}p{2.2cm}p{3.0cm}}
 \hline
 Source name    &T$\rm{_{ex}}$ (K)    & N (cm$^{-2}$)  & N (H$_2$) (cm$^{-2}$)    & Abundance    & Reference\\
 \hline
 Sgr B2        & 20$\sim$60    & $ 0.2 \sim 6\times10^{14}$      & $-$   & $-$         &\citet{turner1975}\\
 Sgr B2(OH)    & 15            & $ 2\times10^{13}$   & $5 \times 10^{23}$  & $4\times 10^{-11}$    &\citet{cummins1986}\\
 Sgr B2(N)     & 500           & $ \sim 2.3\times 10^{14}$       & $3 \times 10^{24}$    & $7.6\times 10^{-11}$  &\citet{nummelin2000}\\
 Sgr B2(N)     & 150           & $5.13(2.05)\times 10^{16}$       & $-$    & $-$  &\citet{belloche2013}\\
 Sgr B2(M)     & 68            & $ 3\times 10^{13}$  & $2 \times 10^{24}$ & $1.5\times 10^{-11}$  &\citet{nummelin2000}\\
 Orion-KL      & 100, 200      & $ 1.1, 0.33\times 10^{16}$        & $-$ & $-$         &\citet{white2003}\\
 NGC 253       & 67         & $ 1.2\times 10^{13}$  & $6.5\times 10^{22}$ & $2\times 10^{-10}$    &\citet{martin2006}\\
 M 82?   & 80.3   & $ 1.2\times 10^{13}$ & $-$   & $-$  &\citet{aladro2011}\\
 G0.253+0.016   & $-$  & $-$    & $1.0\times 10^{22}$   & $-$  &\citet{rathborne2015}\\
 IRAS 20126+4104   & 210 & $ 3.3\times 10^{15}$  & $2.7\times 10^{24}$   & $1.2\times 10^{-9}$   &\citet{palau2017}\\
 IRAS 16293-2422 B     & 100, 300      & $\geq 5, \geq7\times 10^{13}$   & $1.2\times 10^{25}$  & $ \lesssim2\times10^{-10}$  &\citet{coutens2018}\\
 NGC 1333 IRAS2A  & 130    & $ 2.5\times 10^{14}$    & $5\times 10^{24}$ & $\sim 5\times 10^{-11}$    &\citet{coutens2018}\\
 NGC 1333 IRAS2A1   & 150   & $ 8.5\times 10^{14}$   & $5\times 10^{24}$ & $\sim 1.7\times 10^{-10}$    &\citet{belloche2020}\\
 NGC 1333 IRAS4A2  & 150    & $ 8.5\times 10^{14}$   & $5\times 10^{24}$ & $\sim 1.7\times 10^{-10}$    &\citet{belloche2020}\\
 SVS13A    & 220  & $ 2.5\times 10^{15}$    &   & &\citet{belloche2020}\\
 Barnard 1b?  & 60, 200   & $ 1.0, 1.0\times 10^{12}$  
 & $1.4, 1.1\times 10^{25}$  & $7.1, 9.1\times 10^{-14}$  &\citet{marcelino2018}\\
 G+0.693  & 6.3(ortho),\par{6.8(para)}   & $3.8, 27\times 10^{13}$ & $1.35\times 10^{23}$  & $2.3\times 10^{-9}$   & \citet{zeng2018}\\
 NGC 6334I  & 135-285    & $2.3-78\times 10^{15}$   & $-$   & $-$        & \citet{ligterink2020}\\
 \hline
 \multicolumn{5}{l}{Notes. $?$ tentative detection.}
 \end{tabular}
\end{table*}

In the interstellar medium, NH$_2$CN was detected for the first time toward Sgr B2 by \citet{turner1975}, with a column density of 2$\times$10$^{14}$ cm$^{-2}$. Since then, it has been detected in several high-mass star forming regions, Sgr B2(N), Sgr B(M), Orion KL and NGC 6334I \citep{nummelin2000, white2003, ligterink2020}, the extragalactic sources, NGC 253 and M86 \citep{martin2006, aladro2011}, the molecular cluster G0.253+0.016 \citep{rathborne2015}, the massive protostar IRAS 20126+4104 \citep{palau2017}, two solar-type protostars, IRAS 16293-2422B and NGC 1333 IRAS2A \citep{coutens2018}, the young protostar system Barnard 1B \citep{marcelino2018}, and the quiescent giant molecular cloud, G+0.693-0.027 (hereafter G+0.693) \citep{zeng2018}. A list of cyanamide with selected observed information is shown in Table \ref{tab:molecular-column}.

In contrast to NH$_2$CN, HNCNH was only detected through weak maser emission toward the massive star formation region Sgr B2(N) using data from Green Bank Telescope (GBT) PRIMOS survey by \citet{mcguire2012}. The column density of HNCNH was derived to be 2$\times$10$^{13}$cm$^{-2}$, which is about one order of magnitude lower than that of NH$_2$CN. \citet{coutens2018} searched for HNCNH toward 16293-2422B and NGC1333 IRAS2A using ALMA Protostellar Interferometric Line Survey (PILS) and IRAM Plateau de Bure Interferometer (PdBI). However, HNCNH was not found and only a column density upper limit of 3$\times$10$^{15}$cm$^{-2}$ was set. \citet{rivilla2021} also searched for HNCNH in the G+0.693 cloud and set an upper limit of 2.5$\times$10$^{13}$cm$^{-2}$.

In this work, we have added more than 200 reactions relating to NH$_2$CN and HNCNH to the KIDA 2016 network. We consider the experimentally approved or theoretically suggested reactions concerning these two isomers in the models. We proposed new reaction routes which were confirmed by quantum chemistry calculations, and present them in Section 2. We built astrochemical models to simulate the formation and destruction mechanisms of the two prebiotic molecules under the different physical conditions. Also, we explored the possible interstellar environment to form these molecules and illustrate them in Section 3. We discussed the implications of the results in Section 4, and draw conclusions in Section 5.


\section{Chemistry of cyanamide and carbodiimide }
\label{sect:chem}
\subsection{Formation and destruction of cyanamide and carbodiimide}
The formation routes of cyanamide have been explored in many methods. \citet{smith2004} proposed the reaction of CN + NH$_3$ $\rightarrow$ NH$_2$CN + H, but a high barrier was found to form NH$_2$CN and H in gas phase according to a theoretical study \citep{talbi2009}. Later, \citet{blitz2009} experimentally confirmed that this reaction was only the path to produce HCN and NH$_2$. \citet{sleiman2018a} proposed the reaction between CN and CH$_3$NH$_2$ as an efficient route to form NH$_2$CN in ISM through low temperature kinetics experimental and theoretical studies. In their subsequent work, \citet{sleiman2018b} further concluded that the route to form CH$_2$NH$_2$ + HCN is barrierless thus is the major production channel. \citet{puzzarini2020} found CN + CH$_3$NH$_2$ has four reaction channels. \citet{coutens2018} proposed the reaction of NH$_2$ and CN may directly form NH$_2$CN on grain surface.

Carbodiimide may be formed from the isomerization of cyanamide. The activation energy of this process is extremely high in gas phase, while it can be drastically reduced in the presence of water molecules \citep{tordini2003}. \citet{duvernay2004} confirmed that isomerization can become significant with water-ice at low temperatures by experiment. \citet{duvernay2005} further verified through experiments that the isomerization reaction could occur by a photochemical process or when cyanamide is condensed at low temperature (50-140 K) on an amorphous water ice surface or trapped in the water ice, which acts like a catalyst. In addition, \citet{He1991} proposed three gas-phase reactions associated with HNCNH at high temperatures. \citet{yadav2019} obtained two formation routes of HNCNH by quantum chemical calculation in gas phase or on grain surfaces. The first path is the reaction HNC + NH $\rightarrow$ HNCNH. The second one contains two steps: NH + CN $\rightarrow$ HNCN, followed by HNCN + H $\rightarrow$ HNCNH. However, in ISM, the addition reaction would hardly take place in such low-density gas due to the difficulty to release excess energy. Therefore, these addition reactions just occur on dust grains in ISM. 

The chemistry of cyanamide and carbodiimide has been investigated by multiple research groups. But so far, no efficient gas-phase formation reactions have been proposed. The origin of cyanaminde and carbodiimide in the ISM remains unclear. In this work, we ran quantum chemical calculations to evaluate the reactions of the two isomers and their related species. We found that the reaction of NH$_2$ + HNC and HNCN + CH$_2$ to produce NH$_2$CN and HNCNH, respectively, have relative high barriers (Reactions (1)-(2)). Besides, we list some key destruction reactions below (Reactions (3)-(6)).
\protect\\

NH$_2$ + HNC $\rightarrow$ NH$_2$CN + H, \hfill(1)

HNCN + CH$_2$ $\rightarrow$ HNCNH + CH, \hfill(2)

NH$_2$CN + H $\rightarrow$ HNCN + H$_2$, \hfill(3)

HNCNH + H $\rightarrow$ HNCN + H$_2$, \hfill(4)

HNCNH + OH $\rightarrow$ HNCN + H$_2$O,	\hfill(5)

NH$_2$CN + OH $\rightarrow$ HNCN + H$_2$O. \hfill(6)\protect\\

\noindent In addition to the reactions mentioned above, two radicals can likely add together on the grain surface and are assumed to be barrierless (Reaction (7)-(9)), although when the radicals attach to water ice, there might be a barrier \citep{enrique2019}. HNCN can form both cyanamide and carbodiimide through hydrogenation. Therefore, the formation we also propose reactions of HNCN (Reactions (8)-(9)).
\protect\\

HNCN + H $\rightarrow$ NH$_2$CN, \hfill(7)

HNC + N $\rightarrow$ HNCN,  \hfill(8)

NCN + H $\rightarrow$ HNCN. \hfill(9)\protect\\

\noindent Rate coefficients for gas-phase reactions of NH$_2$CN, HNCNH and related species are summarized in Table \ref{tab:gas-phase}. The formulae that are used to calculate rate coefficients are categorized by different reaction types. For cosmic-ray induced photodissociation and UV photodissociation, the formulae are expressed as follows: $k(T)$ = $\alpha \zeta$, $k(T)$ = $\alpha$ exp (-$\gamma$Av) respectively, where $\zeta$ is cosmic-ray ionization rate; Av is the visual extinction (mag). The rate parameters of the cosmic-ray photodissociation and UV photodissociation reactions of cyanamide, carbodiimide and related species (HNCN, NCN) were estimated based on similar reactions already existing in the chemical network. Branching ratios to different product channels are often unmeasured for ion-molecule reactions that can lead to multiple products. To address this, we make the simplifying assumption that the different reaction channels occur equally, and thus have the same branching ratio. And the corresponding rate is calculated as $k(T)$ = $\alpha \beta$ (0.62 + 0.4767 $\gamma (300 / T))$ , except the reactions from \citet{woon2009}, where the corresponding rates are $k(T)$ = $\alpha (T / 300)^{\beta} \exp{(-\gamma / T)}$. This formula is also adopted for the neutral-neutral reactions and dissociative recombination reactions. 

Related surface reactions are listed in Table \ref{tab:activation-energy} with their activation energy barriers. The barriers of free radical groups are assumed to be zero in this work. There is a high isomerization barrier between NH$_2$CN and HNCNH in the gas phase. But on ice surface, this barrier is low enough so that the reaction can take place even at a low temperature \citep{duvernay2005}. Therefore, the surface isomerization reaction is added in the chemical reaction network in Table \ref{tab:activation-energy}. As for the binding energies of NH$_2$CN and HNCNH, we adopt the value of 5556 K for both isomers, as listed in Table \ref{tab:binding-energy}. Besides, the binding energies of all other newly added species are also included in the table. 

\subsection{Quantum chemical calculations}
All quantum chemical calculations reported here were performed with the double-hybrid B2PLYP functions \citep{grimme2006} combined with the may-cc-pVTZ basis set to optimize the structure and to calculate the frequency for all species and transition states involved in the study \citep{dunning1989, papajak2009}. Semi-empirical dispersion contributions were also included by means of the D3BJ model \citep{goerigk2011, grimme2011}. Saddle points were assigned to reaction paths by using intrinsic reaction coordinate (IRC) \citep{fukui1981} calculations at the same B2PLYP level as the identification of reactants and products. To achieve accurate energies for the stationary points from the optimization at the B2PLYPD3/may-cc-pVTZ level, the highly cost-effective method of CCSD(T) \citep{purvis1982, scuseria1988, scuseria1989}, together with the basis sets aug-cc-pVTZ was further used. All quantum chemical calculations were run with the GAUSSIAN 16 program \citep{Gaussian16}.

\clearpage
\onecolumn
\begin{longtable}{p{6cm}p{1.3cm}p{1.3cm}p{1.2cm}p{3.2cm}}
\caption{Summary of the rate coefficients of gas-phase reactions involving cyanamide, carbodiimide and related species.}
 \label{tab:gas-phase}\\
  \hline
  \endfirsthead
 \multicolumn{5}{l}{Continuation of Table \ref{tab:gas-phase}}\\
 \hline
 Reaction    & $\alpha$     & $\beta$        & $\gamma$    & Ref.\\
 \hline
 \endhead
 \hline
 \endfoot
 \endlastfoot

  Reaction   & $\alpha$     & $\beta$        & $\gamma$    & Ref.\\
  \hline
  Cosmic Ray Induced Photodissociation    &(-)   &(-)   &(-)    & \\
  \hline
  NH$_2$CN $\rightarrow$ NH$_2$ + CN                             & 9.50E+03	    & 0	             & 0	       & \citet{harada2010}\\
  NH$_2$CN $\rightarrow$ H + HNCN                                & 9.50E+03	    & 0	             & 0	       & This work\\
  HNCNH $\rightarrow$ NH + HNC                                   & 9.50E+03	    & 0	             & 0	       & This work\\
  HNCNH $\rightarrow$ H + HNCN                                   & 9.50E+03	    & 0	             & 0	       & This work\\
  HNCN $\rightarrow$ N$_2 $ + CH                                 & 9.50E+03	    & 0	             & 0	       & \citet{bise2001}\\
  HNCN $\rightarrow$ HCN + N                                     & 9.50E+03	    & 0	             & 0	       & \citet{bise2001}\\
  HNCN $\rightarrow$ HNC + N                                     & 9.50E+03	    & 0	             & 0	       & \citet{bise2001}\\
  HNCN $\rightarrow$ NCN + H                                     & 9.50E+03	    & 0	             & 0	       & \citet{bise2001}\\
  NCN  $\rightarrow$ N$_2$ + C                                   & 9.50E+03	    & 0	             & 0	       & \citet{bise1999}\\
  NCN  $\rightarrow$ N + CN                                      & 9.50E+03	    & 0	             & 0	       & \citet{bise1999}\\
  \hline
  Photodissociation                                              & $(s^{-1})$       &(-)             &(-)          &\\
  \hline
  NH$_2$CN $\rightarrow$ NH$_2$ + CN                             & 1.00E-09	        & 0	             & 1.7	       & OSU database\\
  NH$_2$CN $\rightarrow$ H + HNCN                                & 1.00E-09	        & 0	             & 1.7	       & This work\\
  HNCNH $\rightarrow$ NH + HNC                                   & 1.00E-09	        & 0	             & 1.7	       & This work\\
  HNCNH $\rightarrow$ H + HNCN                                   & 1.00E-09	        & 0	             & 1.7	       & This work\\
  HNCN $\rightarrow$ N$_2 $ + CH                                 & 1.00E-09	        & 0	             & 1.7	       & \citet{bise2001}\\
  HNCN $\rightarrow$ HCN + N                                     & 1.00E-09	        & 0	             & 1.7	       & \citet{bise2001}\\
  HNCN $\rightarrow$ HNC + N                                     & 1.00E-09	        & 0	             & 1.7	       & \citet{bise2001}\\
  HNCN $\rightarrow$ NCN + H                                     & 1.00E-09	        & 0	             & 1.7	       & \citet{bise2001}\\
  NCN  $\rightarrow$ N$_2$ + C                                   & 1.00E-09	        & 0	             & 1.7	       & \citet{bise1999}\\
  NCN  $\rightarrow$ N + CN                                      & 1.00E-09	        & 0	             & 1.7	       & \citet{bise1999}\\
  \hline
  Ion-neutral      &(-)               &$(cm^3 s^{-1})$ &(-)          &\\
  \hline
  C$^+$ + NH$_2$CN $\rightarrow$  NH$_2$CN$^+$ + C	             & 0.50	            & 1.48E-09	     & 8.57	       & This work\\
  C$^+$ + NH$_2$CN $\rightarrow$  C$_2$N$^+$ + NH$_2$	         & 0.50	            & 1.48E-09	     & 8.57	       & This work\\
  C$^+$ + HNCNH $\rightarrow$  HNCNH$^+$ + C	                 & 0.50	            & 1.57E-09	     & 3.66	       & This work\\
  C$^+$ + HNCNH $\rightarrow$  HNCN + CH$^+$	                 & 0.50	            & 1.57E-09	     & 3.66	       & This work\\
  H$^+$ + NH$_2$CN $\rightarrow$  NH$_2$ + HNC$^+$	             & 0.50	            & 4.73E-09	     & 8.13		   & \citet{woon2009}\\
  H$^+$ + NH$_2$CN $\rightarrow$  NH$_2^+$ + HNC   	             & 0.50	            & 4.73E-09	     & 8.13		   & \citet{woon2009}\\
  H$^+$ + HNCNH $\rightarrow$  HNC$^+$ + NH$_2$    	             & 0.50	            & 4.85E-09	     & 3.66		   & This work\\
  H$^+$ + HNCNH $\rightarrow$  HNC + NH$_2^+$    	             & 0.50	            & 4.85E-09	     & 3.66		   & This work\\
  He$^+$ + NH$_2$CN $\rightarrow$  NH$_2$ + CN$^+$ + He	         & 0.50	            & 2.45E-09	     & 8.13	       & \citet{woon2009}\\
  He$^+$ + NH$_2$CN $\rightarrow$  NH$_2^+$ + CN + He	         & 0.50	            & 2.45E-09	     & 8.13	       & \citet{woon2009}\\
  He$^+$ + HNCNH $\rightarrow$  HNC$^+$ + NH + He	             & 0.50	            & 2.51E-09	     & 3.66	       & This work\\
  He$^+$ + HNCNH $\rightarrow$  HNC + NH$^+$ + He	             & 0.50	            & 2.51E-09	     & 3.66	       & This work\\
  H$_3^+$ + NH$_2$CN $\rightarrow$ NH$_2$CNH$^+$ + H$_2$ 	     & 0.50	            & 2.79E-09	     & 8.13		   & \citet{woon2009}\\
  H$_3^+$ + NH$_2$CN $\rightarrow$ HCN$^+$ + NH$_2$ + H$_2$ 	 & 0.50	            & 2.79E-09	     & 8.13		   & This work\\
  H$_3^+$ + HNCNH $\rightarrow$ NH$_2$CNH$^+$ + H$_2$ 	         & 0.33	            & 2.86E-09	     & 3.66		   & This work\\
  H$_3^+$ + HNCNH $\rightarrow$ HNC$^+$ + NH$_2$ + H$_2$ 	     & 0.33	            & 2.86E-09	     & 3.66		   & This work\\
  H$_3^+$ + HNCNH $\rightarrow$ HNC + NH$_2^+$ + H$_2$ 	         & 0.33	            & 2.86E-09	     & 3.66		   & This work\\
  H$_3$O$^+$ + NH$_2$CN $\rightarrow$ HCN$^+$ + NH$_2$ + H$_2$O  & 0.50	            & 1.25E-09	     & 8.57		   & This work\\
  H$_3$O$^+$ + NH$_2$CN $\rightarrow$ NH$_2$CNH$^+$ + H$_2$O     & 0.50	            & 1.25E-09	     & 8.57		   & This work\\
  H$_3$O$^+$ + HNCNH $\rightarrow$ NH$_2$CNH$^+$ + H$_2$O        & 0.33	            & 1.33E-09	     & 3.66		   & This work\\
  H$_3$O$^+$ + HNCNH $\rightarrow$ HNC$^+$ + NH$_2$ + H$_2$O     & 0.33	            & 1.33E-09	     & 3.66		   & This work\\
  H$_3$O$^+$ + HNCNH $\rightarrow$ HNC + NH$_2^+$ + H$_2$O       & 0.33	            & 1.33E-09	     & 3.66		   & This work\\
  HCO$^+$ + NH$_2$CN $\rightarrow$ NH$_2$CNH$^+$ + CO	         & 0.50	            & 1.13E-09	     & 8.13	       & \citet{woon2009}\\
  HCO$^+$ + NH$_2$CN $\rightarrow$ HCN$^+$ + NH$_2$ + CO	     & 0.50	            & 1.13E-09	     & 8.13	       & This work\\
  HCO$^+$ + HNCNH $\rightarrow$ NH$_2$CNH$^+$ + CO	             & 0.33	            & 1.16E-09	     & 3.66	       & This work\\
  HCO$^+$ + HNCNH $\rightarrow$ HNC$^+$ + NH$_2$ + CO	         & 0.33	            & 1.16E-09	     & 3.66	       & This work\\
  HCO$^+$ + HNCNH $\rightarrow$ HNC + NH$_2^+$ + CO	             & 0.33	            & 1.16E-09	     & 3.66	       & This work\\
  He$^+$ + HNCN $\rightarrow$  CN$^+$ + NH + He	                 & 0.50	            & 2.39E-09	     & 5.30	       & This work\\
  He$^+$ + HNCN $\rightarrow$  CN + NH$^+$ + He	                 & 0.50	            & 2.39E-09	     & 5.30	       & This work\\
  C$^+$ + HNCN $\rightarrow$  HNCN$^+$ + C	                     & 0.50	            & 1.50E-09	     & 5.30	       & This work\\
  C$^+$ + HNCN $\rightarrow$  NCN + CH$^+$	                     & 0.50	            & 1.50E-09	     & 5.30	       & This work\\
  H$^+$ + HNCN $\rightarrow$  NCN$^+$ + H$_2$    	             & 0.50	            & 4.61E-09	     & 5.30		   & This work\\
  H$^+$ + HNCN $\rightarrow$  HNCN$^+$ + H    	                 & 0.50	            & 4.61E-09	     & 5.30		   & This work\\
  H$_3^+$ + HNCN $\rightarrow$ HNCNH$^+$ + H$_2$ 	             & 0.33	            & 2.73E-09	     & 5.30		   & This work\\
  H$_3^+$ + HNCN $\rightarrow$ HNC$^+$ + NH + H$_2$ 	         & 0.33	            & 2.73E-09	     & 5.30		   & This work\\
  H$_3^+$ + HNCN $\rightarrow$ HNC + NH$^+$ + H$_2$ 	         & 0.33	            & 2.73E-09	     & 5.30		   & This work\\
  H$_3$O$^+$ + HNCN $\rightarrow$ HNCNH$^+$ + H$_2$O             & 0.33	            & 1.26E-09	     & 5.30		   & This work\\
  H$_3$O$^+$ + HNCN $\rightarrow$ HNC$^+$ + NH + H$_2$O          & 0.33	            & 1.26E-09	     & 5.30		   & This work\\
  H$_3$O$^+$ + HNCN $\rightarrow$ HNC + NH$^+$ + H$_2$O          & 0.33	            & 1.26E-09	     & 5.30		   & This work\\
  HCO$^+$ + HNCN $\rightarrow$ HNCNH$^+$ + CO	                 & 0.33	            & 1.11E-09	     & 5.30	       & This work\\
  HCO$^+$ + HNCN $\rightarrow$ HNC$^+$ + NH + CO	             & 0.33	            & 1.11E-09	     & 5.30	       & This work\\
  HCO$^+$ + HNCN $\rightarrow$ HNC + NH$^+$ + CO	             & 0.33	            & 1.11E-09	     & 5.30	       & This work\\
  He$^+$ + NCN $\rightarrow$  CN$^+$ + N + He	                 & 0.50	            & 2.27E-09	     & 1.69	       & This work\\
  He$^+$ + NCN $\rightarrow$  CN + N$^+$ + He	                 & 0.50	            & 2.27E-09	     & 1.69	       & This work\\
  C$^+$ + NCN $\rightarrow$  NCN$^+$ + C	                     & 0.50	            & 1.43E-09	     & 1.69	       & This work\\
  C$^+$ + NCN $\rightarrow$  CN + CN$^+$	                     & 0.50	            & 1.43E-09	     & 1.69	       & This work\\
  H$^+$ + NCN $\rightarrow$  NCN$^+$ + H          	             & 0.50	            & 4.39E-09	     & 1.69		   & This work\\
  H$^+$ + NCN $\rightarrow$  CN$^+$ + NH$^+$    	             & 0.50	            & 4.39E-09	     & 1.69		   & This work\\
  H$_3^+$ + NCN $\rightarrow$ HNCN$^+$ + H$_2$ 	                 & 0.33	            & 2.59E-09	     & 1.69		   & This work\\
  H$_3^+$ + NCN $\rightarrow$ CN$^+$ + NH + H$_2$ 	             & 0.33	            & 2.59E-09	     & 1.69		   & This work\\
  H$_3^+$ + NCN $\rightarrow$ CN + NH$^+$ + H$_2$ 	             & 0.33	            & 2.59E-09	     & 1.69		   & This work\\
  H$_3$O$^+$ + NCN $\rightarrow$ HNCN$^+$ + H$_2$O               & 0.33	            & 1.21E-09	     & 1.69		   & This work\\
  H$_3$O$^+$ + NCN $\rightarrow$ CN$^+$ + NH + H$_2$O            & 0.33	            & 1.21E-09	     & 1.69		   & This work\\
  H$_3$O$^+$ + NCN $\rightarrow$ CN + NH$^+$ + H$_2$O            & 0.33	            & 1.21E-09	     & 1.69		   & This work\\
  HCO$^+$ + HNCN $\rightarrow$ HNCN$^+$ + CO	                 & 0.33	            & 1.06E-09	     & 1.69	       & This work\\
  HCO$^+$ + HNCN $\rightarrow$ CN$^+$ + NH + CO	                 & 0.33	            & 1.06E-09	     & 1.69	       & This work\\
  HCO$^+$ + HNCN $\rightarrow$ CN + NH$^+$ + CO	                 & 0.33	            & 1.06E-09	     & 1.69	       & This work\\
  \hline
  Neutral-neutral   & $(cm^3 s^{-1})$  &(-)             &(-)          &\\
  \hline
  NH$_2$ + HNC $\rightarrow$ NH$_2$CN + H  & 8.05E-10	& 0.5	 & 2.69E+03	   & This work\\
  NH$_2$ + CH$_2$ $\rightarrow$ HNCNH + H  & 9.46E-10  & 0.5	 & 5.76E+03	   & This work\\
  HNCNH + H $\rightarrow$ HNCN + H  	   & 1.70E-10  & 0.5	 & 5.03E+03	   & This work\\
  NH$_2$CN + H $\rightarrow$ HNCN + H  	   & 7.75E-10  & 0.5	 & 6.34E+03	   & This work\\
  HNCNH + OH $\rightarrow$ HNCN + H$_2$O   & 8.64E-10  & 0.5	 & 6.62E+02	   & This work\\
  NH$_2$CN + OH $\rightarrow$ HNCN + H$_2$O & 1.41E-11 & 0.0	 & 3.28E+03	   & \citet{espinosa1993}\\
  HNCN + OH $\rightarrow$ HNCN + H$_2$O   & 8.07E-10   & 0.5	 & 6.52E+02	   & This work\\
  HNCN + H $\rightarrow$ HNCN + H$_2$  	  & 1.54E-10   & 0.5	 & 2.29E+03	 & This work\\
  \hline
  Dissociative Recombination  & $(cm^3 s^{-1})$  &(-)   &(-) &\\
  \hline
  NH$_2$CN$^+$ + e$^-$ $\rightarrow$ CN + NH$_2$                 & 1.50E-07	        & -0.5	         & 0	       & This work\\
  NH$_2$CNH$^+$ + e$^-$ $\rightarrow$ HNC + NH$_2$	             & 1.50E-07	        & -0.5	         & 0	       & OSU database\\
  NH$_2$CNH$^+$ + e$^-$ $\rightarrow$ NH$_2$CN + H 	             & 1.50E-07	        & -0.5	         & 0	       & OSU database\\
  NH$_2$CNH$^+$ + e$^-$ $\rightarrow$ HNCNH + H 	             & 1.50E-07	        & -0.5	         & 0	       & This work\\
  HNCNH$^+$ + e$^-$ $\rightarrow$ HNC + NH                       & 1.50E-07	        & -0.5	         & 0	       & This work\\
  HNCNH$^+$ + e$^-$ $\rightarrow$ HNCN + H                       & 1.50E-07	        & -0.5	         & 0	       & This work\\
  HNCN$^+$ + e$^-$ $\rightarrow$ H + NCN                         & 1.50E-07	        & -0.5	         & 0	       & This work\\
  HNCNH$^+$ + e$^-$ $\rightarrow$ HNCN + H                       & 1.50E-07	        & -0.5	         & 0	       & This work\\
  NCN$^+$ + e$^-$ $\rightarrow$ N + CN                           & 1.50E-07	        & -0.5	         & 0	       & This work\\
\hline
\end{longtable}
\clearpage
\twocolumn

\begin{table*}
 \caption{Activation energy barrier values of important surface/mantle reactions.}
 \label{tab:activation-energy}
 \begin{tabular}{p{5.5cm}p{3cm}p{4.5cm}}
  \hline
  Surface reactions  & E$_a$(K)      &Ref.\\
  \hline
  NH$_2$ + HNC $\rightarrow$ NH$_2$CN + H  	 & 2.69E+03	   & This work\\
  HNCN + CH$_2$ $\rightarrow$ HNCNH + CH  	 & 5.76E+03	   & This work\\
  NH$_2$ + CN $\rightarrow$ NH$_2$CN      	 & 0   &\citet{coutens2018}\\
  H + HNCN $\rightarrow$ NH$_2$CN            & 0	       & This work\\
  H + HNCN $\rightarrow$ HNCNH             	 & 0	 &\citet{yadav2019}\\
  NH + HNC $\rightarrow$ HNCNH             	 & 0	 &\citet{yadav2019}\\
  NCN + H$_2$ $\rightarrow$ NH$_2$CN         & 0	       &  This work\\
  H + NCN $\rightarrow$ HNCN                 & 0  & \citet{rivilla2021}\\
  NH + CN $\rightarrow$ HNCN             	 & 0  &\citet{yadav2019} \\
  N + CN $\rightarrow$ NCN                 	 & 0	       & This work\\
  HNCNH + H $\rightarrow$ HNCN + H  	     & 5.03E+03	   & This work\\
  NH$_2$CN + H $\rightarrow$ HNCN + H  	     & 6.34E+03	   & This work\\
  HNCNH + OH $\rightarrow$ HNCN + H$_2$O  	 & 6.64E+02	   & This work\\
  NH$_2$CN + OH $\rightarrow$ HNCN + H$_2$O &3.28E+03 
  &\citet{espinosa1993}\\
  HNCN + OH $\rightarrow$ HNCN + H$_2$O   & 6.54E+02	   & This work\\
  HNCN + H $\rightarrow$ HNCN + H$_2$  	  & 2.29E+03	   & This work\\
  NH$_2$CN $\rightarrow$ HNCNH    & 4.30E+03	 &\citet{duvernay2005}\\
 \hline
 \end{tabular}
\end{table*}

\begin{table}
 \caption{Binding energies of involved species.}
 \label{tab:binding-energy}
 \begin{tabular}{p{2cm}p{2cm}p{2cm}}
  \hline
  Species    & E$\rm{_{des}}$(K)    &Ref.\\
  \hline
  NH$_2$CN  & 5556	 &NH$_2$+CN, \citet{garrod2006}\\
  HNCNH                  & 5556	           & NH$_2$CN\\
  HNCN             	     & 2500	           & NH+CN\\
  NCN                 	 & 2400	           & N+CN\\
\hline
 \end{tabular}
\end{table}

\section{ASTROCHEMICAL MODEL}
\label{sect:model}
To investigate the formation and destruction mechanism of cyanamide and its isomer carbodiimide, we use the three-phase NAUTILUS chemical code \citep{ruaud2016}, which includes the gas, the dust grain surface and the icy mantle. The code allows us to compute abundances of chemical species with respect to time for a given set of physical and chemical parameters by setting up and solving a series of ordinary differential equations. In this study, in addition to the related reactions of cyanamide and carbodiimide, our network also includes updates of the chemistry of HNCO and its metastable isomers \citep{quan2010}, CH$_3$CHNH chemistry \citep{quan2016}, as well as the chemistry of cyanomethanimine \citep{zhang2020}. We have added more than 200 reactions concerning NH$_2$CN and HNCNH. Except for gas-phase reactions  as shown in Table \ref{tab:gas-phase}, the rate constants of reactions are calculated  by the corresponding equations according to \citet{hasegawa1992}, \citet{ruaud2016} on surface and mantle phases. We use the standard oxygen-rich low-metal elemental abundances. The initial abundances \citep{graedel1982, quan2007} are listed in Table \ref{tab:initial-abundance}. All abundances are given with respect to the total hydrogen density. The code simulates chemistry in three phases and also considers various possible exchanges between the different phases via accretion, thermal and non-thermal desorption. The latter includes cosmic-ray desorption, photodesorption via external photons or those via photons produced by cosmic rays \citep{oberg2007}, and reactive desorption using the Rice-Ramsperature-Kessel (RRK) approach \citep{garrod2007}. The surface-mantle and mantle-surface exchange of species is also considered, such as accretion and desorption but with a smaller diffusion rate of the species compared to the processes on the surface \citep{ruaud2016}. The diffusion energy to binding energy ratio is set to be 0.5 for all species. Grains are assumed to be spherical with the radii of 0.1 $\mu$m, The density of dusts is 3 g cm$^{-3}$ and the dust-to-gas mass ratio is 0.01.

\begin{table}
\centering
\caption{Initial elemental abundance of chemical models.}
\label{tab:initial-abundance}
\begin{tabular}{p{2.5cm}p{1.5cm}}
		\hline
		Species         & Abundance\\
		\hline
		He              & $6.00\times10^{-2}$\\
		N               & $2.14\times10^{-5}$\\
		O               & $1.76\times10^{-4}$\\
		H$_{2}$         & $5.00\times10^{-1}$\\
		C$^{+}$         & $7.30\times10^{-5}$\\
		S$^{+}$         & $ 8.00\times10^{-8}$\\
		Si$^{+}$        & $ 8.00\times10^{-9}$\\
		Fe$^{+}$        & $ 3.00\times10^{-9}$\\
		Na$^{+}$        & $ 2.00\times10^{-9}$\\
		Mg$^{+}$        & $ 9.00\times10^{-9}$\\
		P$^{+}$         & $ 3.00\times10^{-9}$\\
		Cl$^{+}$        & $ 4.00\times10^{-9}$\\
		F$^{+}$         & $ 6.69\times10^{-9}$\\
		\hline
\end{tabular}
\end{table}
Cyanamide has been detected in many different types of sources, while carbodiimide is only detected in Sgr B2(N) by maser emission. We classified the sources where cyanamide was detected into three categories: hot cores (e.g. Sgr B2(N), Orion KL) \citep{turner1975, nummelin2000, white2003}, hot corinos (e.g. IRAS 16293-2422B, NGC 1333IRA2A) \citep{coutens2018}, and low-velocity shock regions (e.g. G+0.693, NGC 253) \citep{zeng2018, martin2006}.  Correspondingly, to simulate how NH$_2$CN and HNCNH are formed and destroyed in the ISM, we applied three sets of models with different physical conditions: cold cores, hot cores/hot corinos, and shock models. Regarding G+0.693, a quiescent molecular cloud located within the Sgr B2 star-forming complex in the Central Molecular Zone (CMZ). The shock that is yielded from a cloud-cloud collision \citep{zeng2020}, ultimately sputters icy mantles of grains efficiently. For NGC 253, the origin of the large-scale shocks is also suggested to be associated with a cloud–cloud collision \citep{ellingsen2017}. Furthermore, NGC 253 and G+0.693 have similar physical properties, including gas kinetic temperatures and H$_2$ densities \citep{zeng2020}.

In cold core models, the physical parameters remain constant with a hydrogen density of n$\rm{_H}$ = n(H) + 2n(H$_2$) = 2$\times$10$^{4}$ cm$^{-3}$, the gas and grain temperatures of 10, 15 and 20 K, a visual extinction Av of 10 mag, a cosmic ray ionization rate $\zeta$ of 1.3$\times$10$^{-17}$ s$^{-1}$, UV factor of 1 Habing and a single grain radius of 0.1 $\micron$. 

\begin{table*}
\caption{Physical parameters of hot core/hot corino models}
\label{tab:physical}
\begin{tabular}{p{4.3cm}p{2.2cm}p{1.8cm}p{1.8cm}p{1.0cm}p{2.2cm}}
	\hline
    Source(Stage)  &n$_{\rm H}$(cm$^{-3}$)   &T(K)  &A$_V$ (mag)  &$\zeta$ (s$^{-1}$) & UV factor (Habing)\\
	\hline
	The freefall collapse$^{a,b}$  &3$\times10^{3} \rightarrow$ 1.6$\times10^{7}$ &10               &2 $\rightarrow$ 6.109$\times$10$^{2}$    &1.3$\times$10$^{-17}$ &1\\
	The warm-up$^{b,d}$  &1.6$\times10^{7}$  &10 $\rightarrow$ 150, 200  &6.109$\times$10$^{2}$     &1.3$\times$10$^{-17}$  &1, 10\\
	\hline
    The freefall collapse$^{a,c}$  &3$\times10^{3} \rightarrow$ 6$\times10^{10}$ &10               &2 $\rightarrow$ 1.472$\times$10$^{5}$    &1.3$\times$10$^{-17}$ &1\\
	The warm-up$^{c,d}$  &6$\times10^{10}$  &10 $\rightarrow$ 200, 400  &1.472$\times$10$^{5}$     &1.3$\times$10$^{-17}$  &1\\
	\hline
\multicolumn{5}{l}{Notes. $^a$\citet{garrod2006}, $^b$\citet{bonfand2019}, $^c$\citet{jorgensen2016}, $^d$ \citet{coutens2018}}
\end{tabular}
\end{table*}

According to the different periods of evolution, star formation can be divided into the following two stages: The first is to simulate the growth of the protostar through the accretion of the outer envelope region, namely the freefall collapse process or prestellar phase. During this stage (1 $\times$ 10$^6$ yr), the temperature stays at a constantly low value, the density increases gradually with time, and the extinction also increases. The gas and grain temperatures were set at 10 K, the initial gas density was 3$\times$10$^{3}$ cm$^{-3}$, the final collapse density was set at two values, the low density of 1.6$\times$10$^{7}$ cm$^{-3}$, and the high density of 6$\times$10$^{10}$ cm$^{-3}$. The second stage is the warm-up stage, during which the temperature rises according to the formula $T = T_0 + (T_{max}-T_0)(\Delta t/t_h)^n$ \citep{garrod2006}, where $n$ = 2, t$_h$ = 2 $\times$ 10$^5$ yr, $\Delta t = (t-t_0$), t$_0$ = 1 $\times$ 10$^5$ yr. After an initial cold phase of 1 $\times$ 10$^5$ yr, the gas and grain temperatures increase from 10 K to the maxima T$_{max}$ of 150, 200 and 400 K within a time frame of 2 $\times$ 10$^5$ yr. After reaching the maximum value, the temperature remains unchanged. During this stage, the gas and dust temperatures are assumed to be well coupled. The major physical parameters of hot core/hot corino models are summarized in Table \ref{tab:physical}. 

In the shock models, we use the NAUTILUS code combined with the planar magnetohydrodynamic (MHD) shock code MHD\_VODE \citep{flower2015} to simulate the chemical evolution under shock conditions. As we know, there are no signs of star formation toward G+0.693, of which the large-scale, low-velocity shocks may be yielded from a cloud-cloud collision \citep{zeng2020}. \citet{armijos2020} constrained the cloud-cloud collision age to $\leq$ 0.5 Myr for Sgr B2. \citet{barnes2017} provided a broader time-scale of 0.5–0.9 Myr, assuming that clouds remain quiescent for 0.3–0.5 Myr after star formation. Therefore, before shock coming, we adopted two time-lengths of evolution stages for the cold core condition, 3$\times$10$^{5}$ and 5$\times$10$^{5}$ yr, respectively. The grain temperature of G+0.693 molecular cloud is $\leq$ 30 K \citep{rodriguez2004}, so we assumed that the pre-shock gas and dust temperatures were 20 and 30 K, respectively, while the pre-shock gas density is 2$\times$10$^{4}$ cm$^{-3}$. The cosmic ray ionization rate in G+0.693 is 100 or even 1000 times larger than that in the Galactic disc \citep{zeng2019}. Accordingly, we adopted two higher cosmic ray ionization rates of 1.3$\times$10$^{-15}$, and 1.3$\times$10$^{-14}$ s$^{-1}$. In this study, we have considered weak C-type shock waves with a speed of 10 km s$^{-1}$. The major physical parameters of the shock wave models are summarized in Table \ref{tab:shock}. 

\begin{table}
 \caption{Physical parameters of C-type shock models.}
 \label{tab:shock}
 \centering
 \begin{tabular}{cc}
  \hline
   Parameter	                                      & Value\\
  \hline
   Initial density (cm$^{-3}$) $n_{\rm H}$                   & 2$\times$10$^{4}$\\
   Pre-shock temperature (K)                                 & 20, 30\\
   Radiation field (Habing) UV factor                        & 1\\
   Cosmic ray ionization rate~(s$^{-1}$) $\zeta$             & 1.3$\times$10$^{-15}$, 1.3$\times$10$^{-14}$\\
   Visual extinction (mag)  $A_{\rm V}$                      & 10 \\
   Magnetic field strength parameter  $b$	             & 1.0\\
   Shock speed (km s$^{-1}$) $u{\rm _s}$	             & 10\\
  \hline
\end{tabular}
\end{table}

\section{RESULTS}
\subsection{Cold cores}
\begin{figure*}
\includegraphics[scale=0.7]{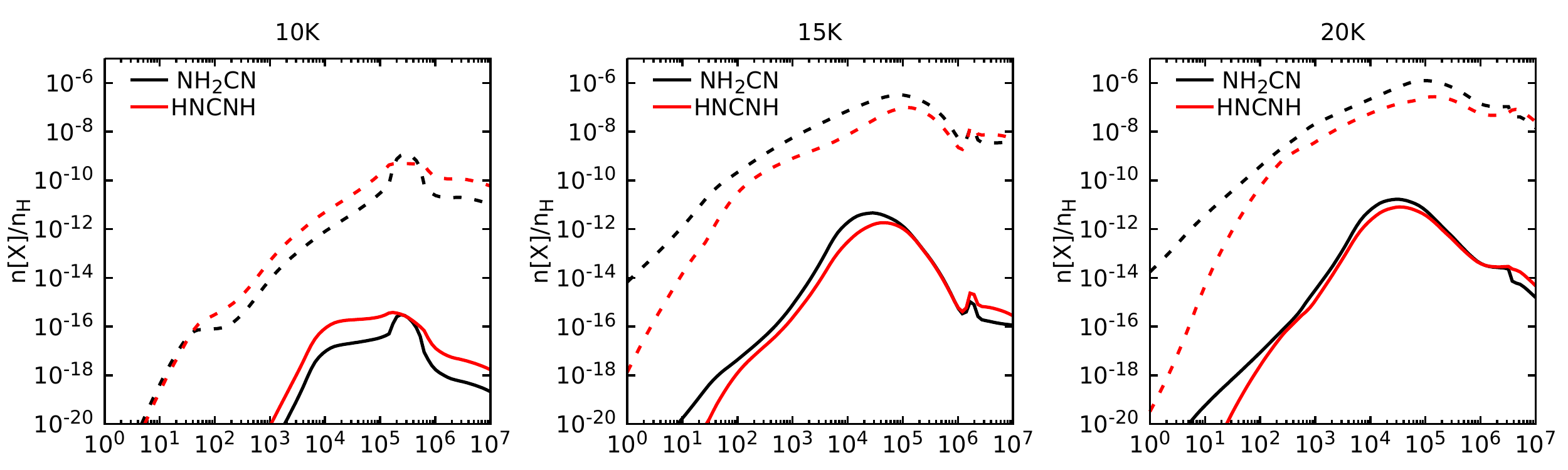}
\caption{The calculated abundances of NH$_2$CN and HNCNH in the gas phase and on grain mantles, including surface and icy mantle, are plotted versus time for cold core models at temperatures of 10, 15, and 20 K. Solid lines correspond to the gas-phase molecules, dotted lines indicate the same species on grains.}
\label{fig:coldCore}
\end{figure*}

 Results from cold core models are shown in Fig. \ref{fig:coldCore}. It can be seen that cyanamide and carbodiimide are hardly produced in the gas phase, but can be slowly formed on the grain mantles at 10 K. The abundances of cyanamide and carbodiimide reach 10$^{-9}$ and 10$^{-10}$ on the grain at the time of $\sim10^5$ yr, respectively. In those models, carbodiimide is mainly produced by the isomerization reaction of cyanamide on the grains. This conclusion agrees with the experiment by \citet{duvernay2005}. In the experiment, cyanamide formed carbodiimide when it was on amorphous water ice surfaces or was trapped in the water ice where the water ice acted like a catalyst and greatly lower the isomerization barrier. Therefore, cyanamide's abundance becomes lower than carbodiimide's at the time of $\sim$30 yr. At 15 and 20 K, gaseous cyanamide isomers’ abundance peaks are as low as $\leq$ 10$^{-11}$, suggesting that they are still ineffectively produced. After 100 years, the cyanamide isomers are produced in a great amount on grain mantles. The maximum fractional abundances of cyanamide and carbodiimide can reach to 10$^{-7}$ and 10$^{-8}$ at 15 K, and 10$^{-6}$ and 10$^{-7}$ at 20 K, respectively. Under these conditions, the main formation routes of cyanamide are the addition reactions of free radicals on the grain surface, NCN + H$_2$, HNCN + H and NH$_2$ + CN. For carbodiimide, the major formation reaction is hydrogenation of HNCN. In these processes, even at the low temperatures, small molecules still can overcome the diffusion barriers, and react to form large molecules like NH$_2$CN and HNCNH on the grain surface.

\subsection{Hot cores/hot corinos}
\begin{figure*}
\includegraphics[scale=0.6]{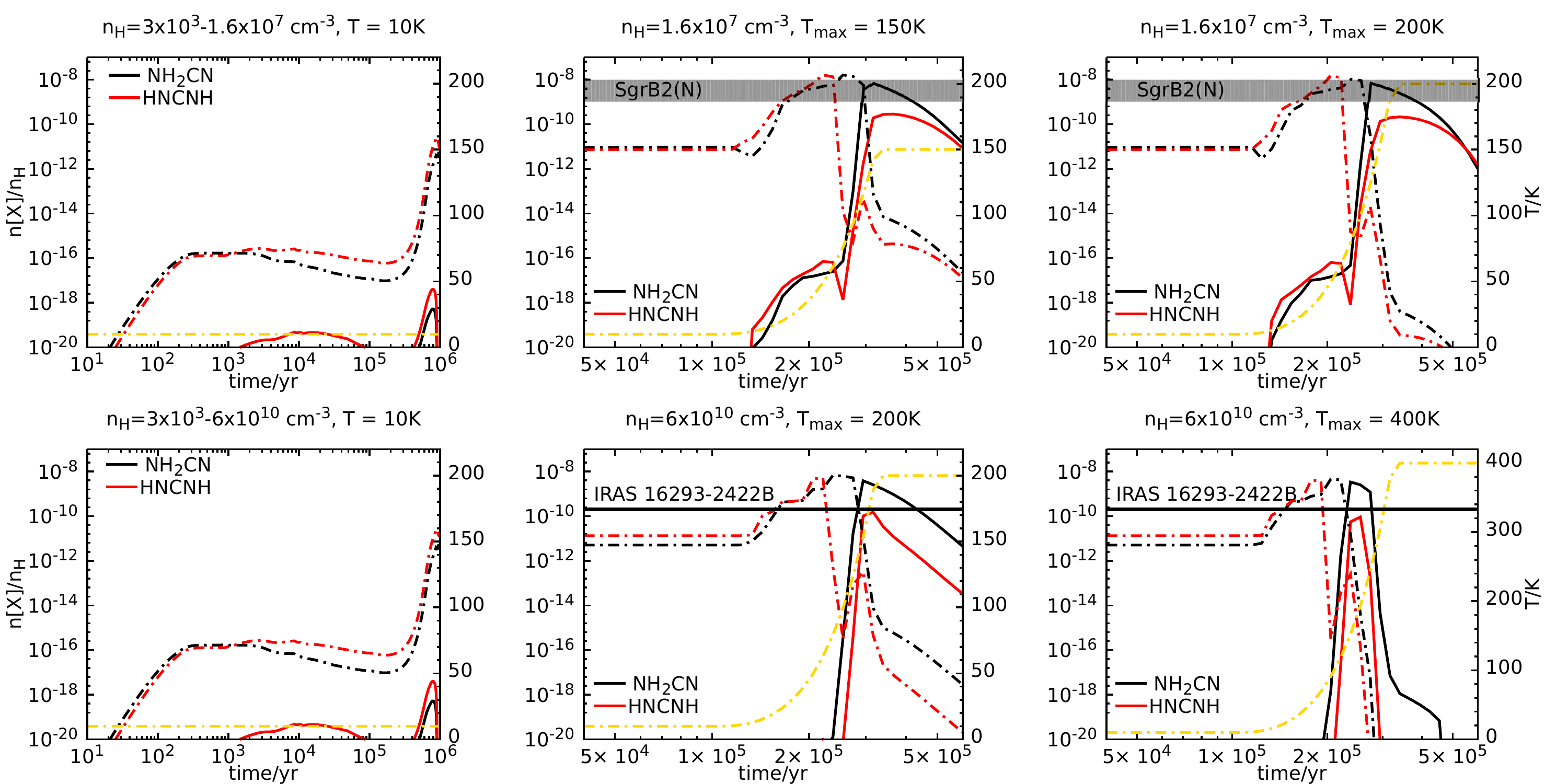}
\caption{The calculated abundances of NH$_2$CN and HNCNH in the gas phase and on grain mantles, including surface and icy mantle, are plotted versus time for freefall collapse and warm-up stages in hot core/hot corino models. During the freefall collapse stage, the initial density is 3$\times$10$^{3}$ cm$^{-3}$ in both left two panels, while the final collapse densities are 6$\times$10$^{10}$ cm$^{-3}$ and 1.6$\times$10$^{7}$ cm$^{-3}$ respectively, and the temperatures remain at a constant value of 10 K. During the warm-up stage, T$_{max}$ adopt two values, 150 and 200 K for upper panels, and 200 and 400 K for bottom panels. Solid lines correspond to the gas-phase abundances, dotted lines indicate abundances on the grains. The gray rectangle represents the observed abundance with $\pm$ a factor of 3 uncertainty for NH$_2$CN in Sgr B2(N), while the horizontal lines are showing upper limit of NH$_2$CN abundances toward IRAS 16293-2422B. Golden dotted lines denote temperature profiles.}
\label{fig:prestellar}
\end{figure*}

Results of hot core / hot corino models are shown In in Fig. \ref{fig:prestellar}. The leftmost panels show the variations of calculated abundances of NH$_2$CN and HNCNH versus time during the freefall collapse stage. In the gas phase, the two species are produced with very low abundances in two collapse densities, while they can be produced in a large amount after 4$\times$10$^{5}$ yr on grains with the rapidly increasing density. The main formation route is the reaction of HNCN + H on grain surface and icy mantle for both NH$_2$CN and HNCNH. Besides, NH$_2$CN can also be formed by the reaction of NCN + H$_2$. The maximum fractional abundance reaches 10$^{-11}$. We found that the two different collapse densities affect little on the fractional abundances of the isomers in gas-phase nor on grain surface. 
 
The right two columns of Fig. \ref{fig:prestellar} show computational results from four models, corresponding to different maximum temperatures in warm-up stages and different gas densities gained from the freefall collapse conditions. As shown in the two top right panels, the warm-up models can produce sufficient abundances of NH$_2$CN and HNCNH during the warm-up to maximum temperatures of 150 and 200 K with the hydrogen density of 1.6$\times$10$^{7}$ cm$^{-3}$. Their peak fractional abundances can reach 6$\times$10$^{-9}$ and 2$\times$10$^{-10}$ at T$_{max}$ of 150 and 200 K, respectively. The physical parameters of top panels represent the sources of Sgr B2(N) suggested by \citet{bonfand2019}. The observational result is taken from \citet{belloche2013}. The emission of NH$_2$CN corresponding to the warm component is optically thin for a source size of 2$^{''}$, the rotational temperature is 150 K, the column density is 5.13$\times$10$^{16}$ cm$^{-2}$. In the other set of parameters, the density is 10$^{13}$ cm$^{-2}$ for the size of 60$^{''}$ where the gas is diluted. We adopt the hydrogen column density of 1.54$\times$10$^{25}$ cm$^{-2}$ for the hot cores \citep{bonfand2017}. The observed abundance of cyanamide is $\sim$ 3$\times$10$^{-9}$. We conclude that our simulated results are in good agreement with the observation within the time range of 2.9-4.2$\times$10$^{5}$ yr. Carbodiimide has been detected in Sgr B2(N) \citep{mcguire2012}. Its column density was anticipated to be $\sim$2$\times$10$^{13}$ cm$^{-2}$ according to \citet{duvernay2005}, where the HNCNH abundance is $\sim$10$\%$ of the abundance of NH$_2$CN from ice experiment. In our models, the calculated ratio of HNCNH to NH$_2$CN is $\sim$9.4$\%$ at the time of 3.65$\times$10$^{5}$ yr when HNCNH abundance reaches its peak for T$_{max}$ = 150 K. Similarly, the ratio is $\sim$8.6$\%$ at the time of 3.40$\times$10$^{5}$ yr for T$_{max}$ = 200 K. The simulated ratios are very close to the experimental value of 10$\%$. However, when NH$_2$CN abundance reaches its peak value, the ratio is getting lower. In the gas phase, the isomers mainly come from the surface by thermal desorption processes. A large amount of surface NH$_2$CN and HNCNH desorb into the gas phase at temperature up to $\sim$140 K. This agrees with the experimental results by \citet{duvernay2005}, in which the desorption temperature of HNCNH is determined to be 130-140 K. We used the same binding energy of 5556 K for both isomers, as shown in Table \ref{tab:binding-energy}. From Fig \ref{fig:prestellar}, it can be seen that when the temperature increases, the two isomers' abundances rise significantly. This is because the thermal energy is sufficient to drive reactants to overcome the diffusion barriers on grain surface at $\sim$50 K. During the warm-up stage, the major formation reactions are NCN + H$_2$ for NH$_2$CN, HNC + NH for HNCNH on the grain surface. Besides, the isomers are destroyed by free radical OH on the grain surface, especially carbodiimide is. Eventually, when the temperature reaches 70 K, the isomers become to desorb efficiently, leading to their peak abundances at 140 K in the gas phase. During the warm-up stage, the two isomers desorb efficiently from the dust surface into the gas phase by thermal desorption. This agrees with the mechanism introduced by \citet{garrod2006} to study interstellar complex organic molecules. After reaching the peak, the abundances of the two isomers subsequently decrease due to active destruction reactions in the gas-phase, including reactions with positive ions such as H$^+$, H$_3^+$, C$^+$, He$^+$ and H$_3$O$^+$ and the free radical OH.

As shown in the two bottom right panels of Fig. \ref{fig:prestellar}, the warm-up models overproduce NH$_2$CN and HNCNH during the warm-up to T$_{max}$ of 200 and 400 K with the hydrogen density of 6$\times$10$^{10}$ cm$^{-3}$. The peak fractional abundances reach 3$\times$10$^{-9}$ for both NH$_2$CN and HNCNH. The physical parameters of bottom panels correspond to the source of IRAS 16293-2422B \citep{jorgensen2016, coutens2018}. The observational results originated from \citet{coutens2018} and the abundance of cyanamide is derived to be $\leq$ 2$\times$10$^{-10}$. Therefore, our simulated abundances are larger than the observation at the time range of 2.9-4.2$\times$10$^{5}$ yr and 2.3-2.7$\times$10$^{5}$ yr. Under the conditions of T$_{max}$ reaches 200 and 400 K, cyanamide and carbodiimide also firstly form on the grain surface, then desorb into gas phase by thermal process when temperature increases to 70 K. The main formation reactions are the same as those described  above in the case of low gas density. On the other hand, the main destruction reactions are ion-neutral reactions by H$_3^+$ for T$_{max}$ = 200 K. Besides, there is a rapid decline of the fractional abundance of NH$_2$CN and HNCNH at T$_{max}$ of 400 K. That is because at this high temperature not only the rates of formation reactions increase but destruction reactions are also faster. The latter include ion-neutral reactions by H$_3^+$, and by atomic H. Although these reactions have high barriers, they can still take place at a high temperature above 200 K. The simulated ratio of NH$_2$CN to HNCNH is $\sim$6$\%$ when HNCNH abundance reaches its peak value for T$_{max}$ of 200 K, while the ratio is $\sim$4$\%$ for T$_{max}$ of 400 K. The cause is that the reaction barrier of HNCNH + H is smaller than that of NH$_2$CN + H, as shown in Table \ref{tab:activation-energy}. The reaction rate of the former is higher than the latter. HNCNH is more quickly destroyed than NH$_2$CN is.

\begin{figure*}
\includegraphics[scale=0.7]{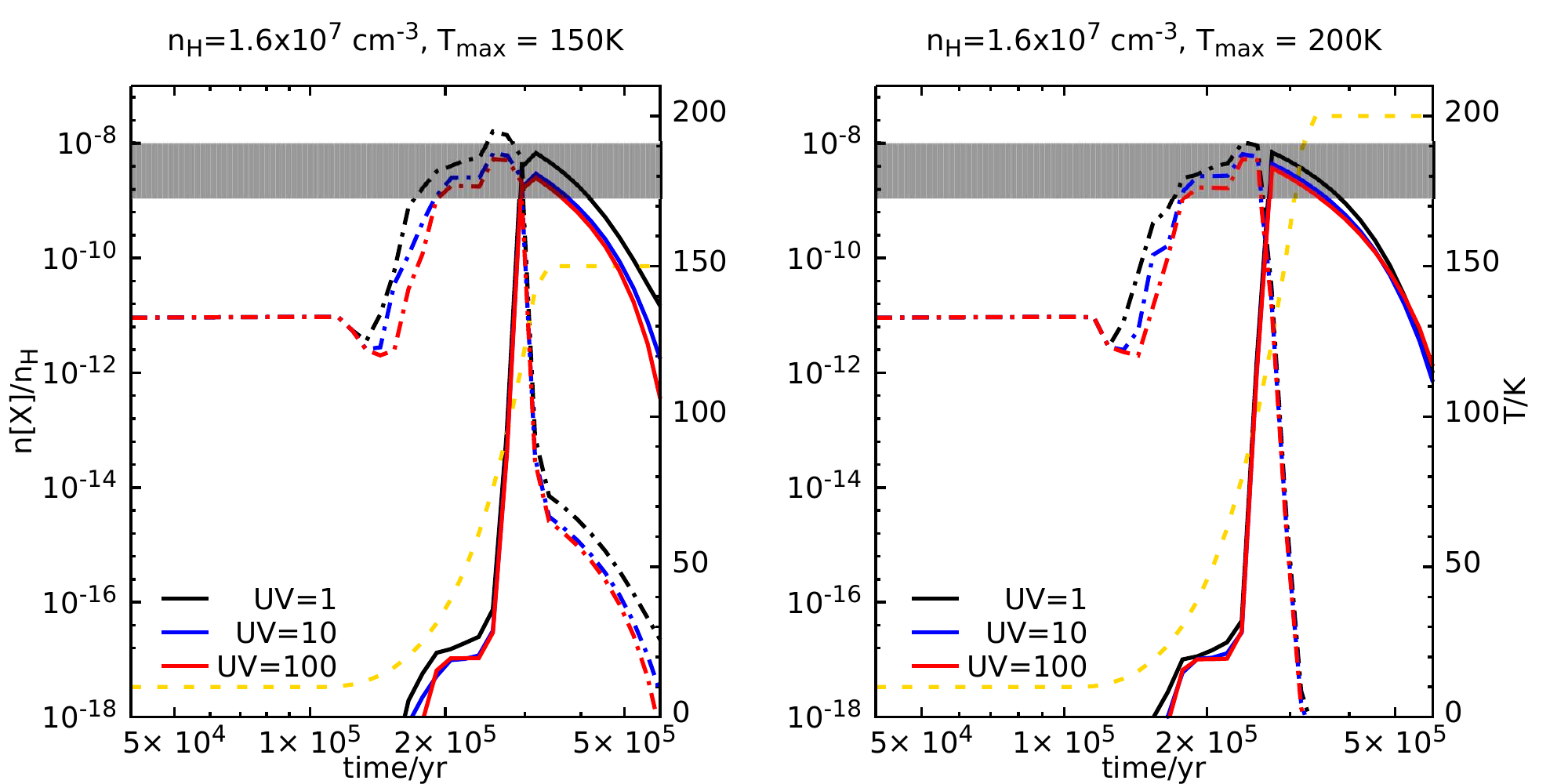}
\caption{The calculated abundances in the gas phase and on the grain mantle, are plotted versus time for warm-up models with the gas density of 1.6 $\times$10$^{7}$ cm$^{-3}$. The left panel denotes T$_{max}$ of 150 K with different UV factors, and the right panel responds T$_{max}$ of 200 K. Only cyanamide is shown as an example here, and carbodiimide has similar tendency. Solid lines correspond to the gas-phase abundances, dotted lines indicate abundances on the grains. The gray rectangle represents the observed abundance with $\pm$ a factor of 3 uncertainty for NH$_2$CN in Sgr B2(N). Golden dotted lines denote temperature profiles.}
\label{fig:prestellar_uv}
\end{figure*}

There is high UV radiation in high mass star formation regions. Therefore, we used two higher values of UV factor, 10 and 100 times larger than the normal value for models with the hydrogen density of 1.6 $\times$10$^{7}$ cm$^{-3}$ to further simulate the physical conditions in Sgr B2(N). In the models, only the warm-up stage adopts the high UV factor. The results are shown in Fig. \ref{fig:prestellar_uv}, It can be seen that fractional abundances of cyanamide are lower comparing to the results from the standard UV factor computations for both T$_{max}$ of 150 and 200 K. Stronger UV radiations destroy the two isomers at higher chemical reaction rates. The peak abundance of cyanamide drops from 6.79$\times$10$^{-9}$ to 2.98$\times$10$^{-9}$ at the time of 3.16$\times$10$^{5}$ yr as the UV factor increases from 1 to 10 for T$_{max}$ of 150 K. For the case of UV factor of 100, it has the same trend as the UV factor of 10, and the peak abundance of cyanamide deduced to 2.53$\times$10$^{-9}$ at the same time. For T$_{max}$ of 200 K, the abundance of cyanamide reduced from 6.91$\times$10$^{-9}$ to 4.37$\times$10$^{-9}$ at the time of 2.74$\times$10$^{5}$ yr when the UV factor increases from 1 to 10. With UV factor 100, the abundance of cyanamide further reduced to 3.74$\times$10$^{-9}$. Fig. \ref{fig:prestellar_uv} shows that the UV factor affects more on T$_{max}$ = 150 K than on T$_{max}$ = 200 K. The effects of UV factors 10 and 100 have a small difference on cyanamide's abundances. The simulated cyanamide abundances also agree well with the observational result within a certain time range, but it lasts a shorter period than using the standard UV factor. Although it is not shown here, the high UV factor has the same effect on the fractional abundance of carbodiimide as on that of cyanamide.

\begin{figure*}
\includegraphics[scale=0.7]{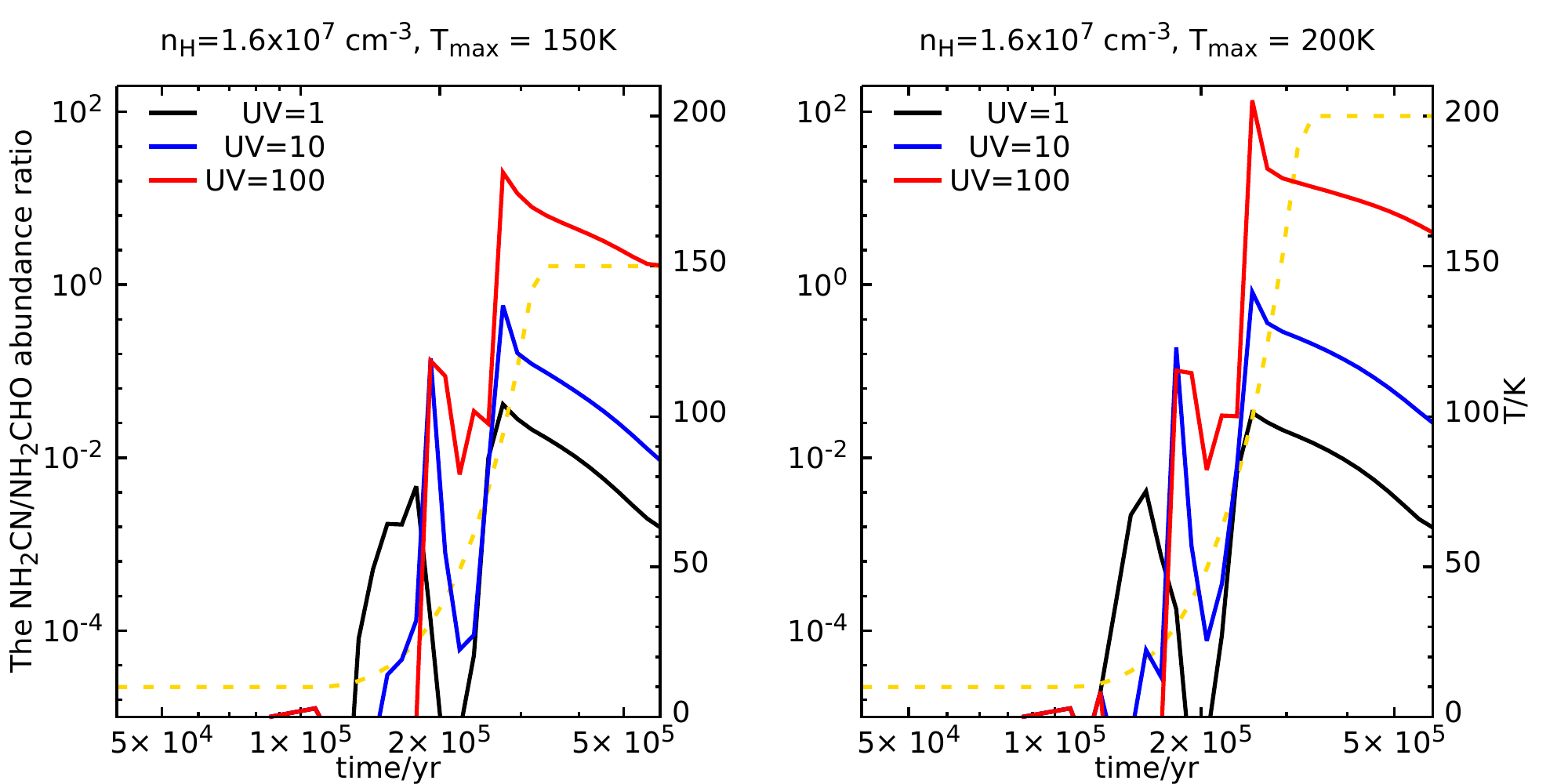}
\caption{The calculated abundance ratios between the NH$_2$CN and NH$_2$CHO in the gas phase are plotted versus time for warm-up models with the gas density of 1.6 $\times$10$^{7}$ cm$^{-3}$. The left panel denotes T$_{max}$ of 150 K with different UV factors, and the right panel responds T$_{max}$ of 200 K. Golden dotted lines denote temperature profiles.}
\label{fig:nh2cn_nh2cho}
\end{figure*}

Formamide (NH$_2$CHO) is the simplest molecule which includes peptide bond -NH-C(=O)-. It has been observed in a wide variety of sources including not only hot cores, hot corinos, protostellar shocked regions, but also comets. Detailed observational results are summarized in Table 1 of \citet{lopez2019}. On the other hand, NH$_2$CN hasn't been detected in any comets so far. It was only detected in hot and dense regions like SgrB2(N), IRAS 16293-2422B, and the source in which wide low-velocity shock waves exist, such as G+0.693. The ratio of NH$_2$CHO to NH$_2$CN varies in different types of sources \citep{coutens2018}. In our study, NH$_2$CHO has the same desorption energy with NH$_2$CN (5556 K), which is close to the desorption energy of water, so that they have the same desorption temperature. NH$_2$CHO is mainly produced by the reaction of NH$_2$ + H$_2$CO $\rightarrow$ NH$_2$CHO + H in the gas phase \citep{garrod2008}. We use the rate coefficient of NH$_2$ + H$_2$CO in the gas phase from \citet{skouteris2017}’s computations, and add the reaction on grain surface as well. NH$_2$CN is mainly produced by surface reactions of HNCN + H, NCN + H$_2$ and NH$_2$ + CN. NH$_2$CHO and NH$_2$CN have the competition to use NH$_2$ as the precursor on the grain surface. The formation reaction of formamide has a low potential barrier (4.48 K) so that it is relatively easier to form than cyanamide is. As the result, NH$_2$CHO exists more widely than NH$_2$CN in ISM. As shown in Fig. \ref{fig:nh2cn_nh2cho}, in standard UV factor models, the calculated ratio of NH$_2$CN to NH$_2$CHO is 0.02 at the time of 3.16$\times$10$^{5}$ yr in which both of them reach their peak abundance positions for T$_{max}$ of 150 K. Similarly, the ratio is 0.03 at the time of 2.74$\times$10$^{5}$ yr for T$_{max}$ of 200 K. The simulated ratios agree well with the observational value of $\sim$0.02-0.04 toward Sgr B2(N) \citep{coutens2018}. In UV factor 10 models, the ratio of NH$_2$CN to NH$_2$CHO increases from 0.02 to 0.12 and from 0.03 to 0.36 for T$_{max}$ of 150 K and 200 K, respectively. In UV factor = 100 models, the ratio is greater than 1. Therefore, the ratio increases with higher UV factor. A strong UV radiation inhibits the formation of NH$_2$CHO. Under strong UV radiation conditions, more NH$_2$ are destroyed by H, which are produced by photodissociation reactions. Formation efficiency of NH$_2$CHO thus decreases. While it affects less on NH$_2$CN, the latter has more formation routes other than from the reactant NH$_2$.

\begin{figure*}
\includegraphics[scale=0.7]{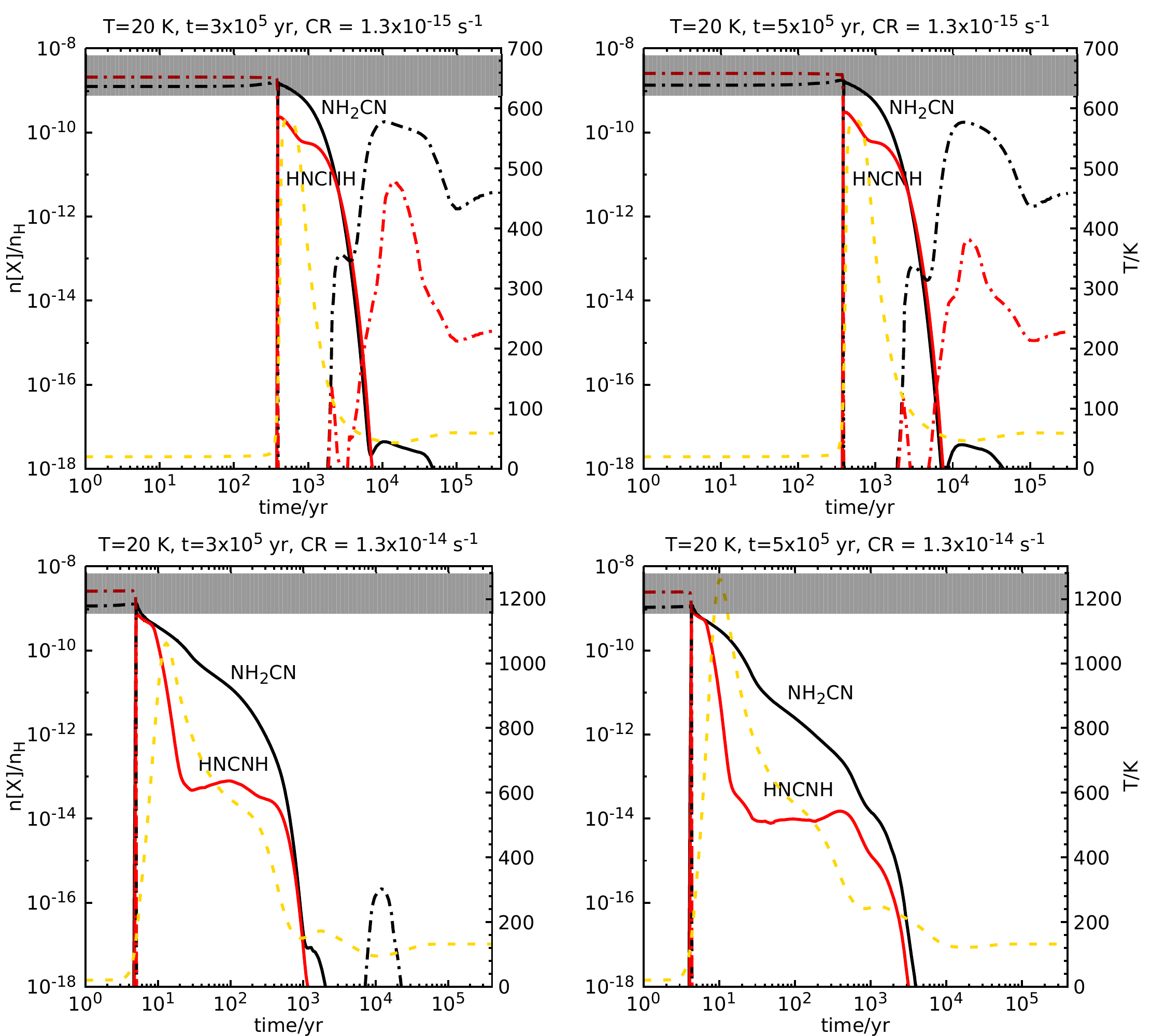}
\caption{The calculated abundances of NH$_2$CN and HNCNH in the gas phase and on grain mantles, including surface and icy mantle for our C-type shock models. The results are plotted versus time after cold core models with the evolution time of 3, 5$\times$10$^{5}$ yr at pre-shock temperatures of 20 K with higher $\zeta$. Solid lines correspond to the gas-phase abundances, and dotted lines indicate abundances on the grains. The gray rectangle represents the observed abundance with $\pm$ a factor of 3 uncertainty for NH$_2$CN in G+0.693. Golden dotted lines denote temperature profiles.}
\label{fig:s20kCR}
\end{figure*}

\subsection{C-type shocked regions}
Results of C-type shock models are shown in Fig. \ref{fig:s20kCR}. In the pre-shock stage, cyanamide and its isomer, carbodiimide, have been produced in a large amount on the grain surface, while the gas phase formation reactions are inefficient because of the low temperature. After shock comes at the velocity of $\sim$10 km s$^{-1}$, the gas density is increased by an order of magnitude, eventually leading to the final density of 10$^5$ cm$^{-3}$. The gas and dust temperatures rapidly increase to over five hundred Kelvin. In Fig. \ref{fig:s20kCR}, t = 0 denotes the onset of the shock. At this time, cyanamide and carbodiimide abundances are taken from the exit values of the previous stage of the cold core models. Then the gaseous cyanamide and carbodiimide fractional abundances dramatically increase with the time when shock propagates. Meanwhile, grain surface abundances of these molecules decrease rapidly. The shock heats the grain and sputters molecules from the grain mantles. Comparing to the observation, all of these abundances can reach the range of the observation, but within different time periods. For $\zeta$= 1.3$\times$10$^{-15}$s$^{-1}$, the range of time is 3.9-7.6$\times$10$^{2}$ yr after 3$\times$10$^{5}$ yr of cold core conditions. This range is shorter (3.9-6.8$\times$10$^{2}$ yr) if the cold core evolves for 5$\times$10$^{5}$ yr, as shown in the top panels of Fig. \ref{fig:s20kCR}. For $\zeta$= 1.3$\times$10$^{-14}$s$^{-1}$, there are only a small time-window in which the result agrees with the observation, which are 50 and 40 yr with the evolution of time t = 3$\times$10$^{5}$ and 5$\times$10$^{5}$ yr for cold core conditions, respectively, as shown in the bottom panels of Fig. \ref{fig:s20kCR}. We found that molecular peak values can be reached at a shorter time with higher cosmic ray ionization rates. That is because stronger cosmic ray accelerates the formation reactions of free radical so that lager molecules form more quickly on grain surface. Using high cosmic ray ionization rates reduces the time to reach peak abundances. Besides, the two isomers' abundances reach the range of observation on the grain surface before the shock comes under cold core conditions with the evolution of time t = 3$\times$10$^{5}$ and 5$\times$10$^{5}$ yr. As the result, their gas phase abundances may agree with the observation after shock sputters them off the dust grains. The final temperature varies with the value of $\zeta$. In addition, when $\zeta$ is 1.3$\times$10$^{-15}$ s$^{-1}$, the past-shock temperature stays at $\leq$ 65 K. Therefore, small species accrete back on the grain from the gas phase, and complex organic molecules (COMs) will form on grain surface, like cyanamide and carbodiimide, as shown in top panels of Fig. \ref{fig:s20kCR}. When $\zeta$ is 1.3$\times$10$^{-14}$ s$^{-1}$, all species have been sputtered by shock into gas phase at peak temperatures (above 1000 K). Then the temperature starts to drop, still above 130 K. Almost all cyanamide molecules no longer exist on grain as the temperature is equal to its' desorption temperature. Molecules that are smaller than cyanamide are hard to deplete on grain. So, there are no COMs like cyanamide could form on grain surfaces. Two examples are shown in bottom panels of Fig. \ref{fig:s20kCR}. For pre-shock temperature of 30K, our results agree with the observation using $\zeta$ = 1.3$\times$10$^{-14}$ s$^{-1}$, but this is not for the case for $\zeta$ = 1.3$\times$10$^{-15}$ s$^{-1}$. In the shock models, the origin of cyanamide and carbodiimide is also from the grain surface by free radical reactions, which are NH$_2$ + CN, HNCN + H and NCN + H$_2$ for cyanamide, and HNCN + H for carbodiimide. This is similar to results under cold core conditions. So, we speculated that the two isomers are mainly formed by sublimation process to release COMs, which were previously formed in the solid phase, into the gas phase. The destruction reactions include cosmic ray induced photodissociation, and ion-neutral reactions by H$_3^+$ , H$_3$O$^+$ for $\zeta$ of 1.3$\times$10$^{-15}$s$^{-1}$. When $\zeta$ is 1.3$\times$10$^{-14}$s$^{-1}$, the destruction reactions are ion-neutral reactions by H$^+$ and the reaction of NH$_2$CN/HNCNH + H. According to our calculations, a strong cosmic ray ionization ($\zeta \geq$10$^{-15}$ s$^{-1}$), shortens the time for the isomers to reach their peak abundances, and decreases the efficiency of their formation on dust grains. 

As a precursor of cyanamide and carbodiimide, HNCN also has been detected toward G+0.693 with the abundance of 9.1$\times$10$^{-11}$, which is an order of magnitude lower than cyanamide abundance. Our simulated results reach $\sim$10$^{-9}$, one order of magnitude larger than the observational result, but the high abundance of HNCN only lasts a very short period of time because HNCN is rapidly converted into cyanamide and carbodiimide by hydrogenation. 

According to \citet{merz2014}, NH$_2$CN, HNCNH and C$_3$NH may create a tautomer of adenine with retrosynthetic analysis. The three molecules are all detected in ISM, but are in different types of sources. So far, C$_3$NH was only detected in a standard cold core TMC-1, cyanamide was mainly detected in hot cores/corinos or shock regions, and carbodiimide was only detected in the hot core Sgr B2(N). Therefore, we preliminarily infer that they are produced under different physical conditions, and it is very unlikely that they could appear in the same region. C$_3$NH's fractional abundance is $\sim$2$\times$10$^{-11}$ from our shock models. This agrees with the suggested value of 1.5$\times$10$^{-11}$ in the tentative detection by \citet{zeng2018}. Besides, the three-body reaction to form the tautomer is hard to occur in ISM.

\section{CONCLUSIONS}
In this work, we ran chemical simulations to calculate the evolution of NH$_2$CN and HNCNH in ISM. The main results are summarized as follows:

1. Cyanamide has been formed during the prestellar phase in hot cores or hot corinos. The formation process is interpreted as resulting from the sublimation of the icy mantles of dust grains. This agrees with the observations that this molecule generally has been found in the compact and hot regions with T$\ge$100 K. A similar process could also occur through shocks, like in the region of G+0.693 where cyanamide was detected. 

2. Our modelling results suggest that cyanamide and carbodiimide molecules come from surface chemistry formation at early evolutionary stages. They are subsequently released back to the gas phase, either by thermal process(in hot cores or hot corinos) or by shock-induced desorption (in shock regions). Comparing with the observations, our simulations are well verified. We found that cyanamide and carbodiimide are mainly formed by free radical reactions on dust grain surfaces, and are destroyed by positive ions in gas phase, as well as OH and H both in gas phase and on grain surfaces. The two isomers are difficult to form in gas phase because of high barriers of gaseous formation reactions. Thus, they are unlikely to be detected in the cold dark cloud cores. 

3. Strong UV radiation significantly lower the abundances of cyanamide and carbodiimide. Stronger UV destroys the two isomers with faster chemical reaction rates. Besides, from high UV factor models, the ratio of NH$_2$CN to NH$_2$CHO increases comparing to a standard UV factor.

4. Strong cosmic ray ionization shortens the time for the isomers to reach peak abundances, and decreases the efficiency of their formation on dust grains.

5. C$_3$NH is formed by gas reactions at low temperatures, and its fractional abundance is low in our simulations where cyanamide has relatively high abundances. Therefore, it is an inefficient route to form a tautomer of adenine by starting from molecules C$_3$NH, HNCNH and NH$_2$CN in ISM.

\section*{Acknowledgements}

This work was supported by the Natural Science Foundation of Xinjiang Uygur Autonomous Region (2022D01A156), the National Key R\&D Program of China (No.2022YFA1603103), the "Tianchi Doctoral Program 2021", the National Natural Science Foundation of China under grant 12203091, 11973075, 12173075.

\section*{DATA AVAILABILITY STATEMENT}
The datasets generated during this study are available in the article and from the corresponding author upon request.



\bibliographystyle{mnras}
\bibliography{bibtex} 

\begin{thebibliography}{}
\makeatletter
\relax
\def\mn@urlcharsother{\let\do\@makeother \do\$\do\&\do\#\do\^\do\_\do\%\do\~}
\def\mn@doi{\begingroup\mn@urlcharsother \@ifnextchar [ {\mn@doi@}
  {\mn@doi@[]}}
\def\mn@doi@[#1]#2{\def\@tempa{#1}\ifx\@tempa\@empty \href
  {http://dx.doi.org/#2} {doi:#2}\else \href {http://dx.doi.org/#2} {#1}\fi
  \endgroup}
\def\mn@eprint#1#2{\mn@eprint@#1:#2::\@nil}
\def\mn@eprint@arXiv#1{\href {http://arxiv.org/abs/#1} {{\tt arXiv:#1}}}
\def\mn@eprint@dblp#1{\href {http://dblp.uni-trier.de/rec/bibtex/#1.xml}
  {dblp:#1}}
\def\mn@eprint@#1:#2:#3:#4\@nil{\def\@tempa {#1}\def\@tempb {#2}\def\@tempc
  {#3}\ifx \@tempc \@empty \let \@tempc \@tempb \let \@tempb \@tempa \fi \ifx
  \@tempb \@empty \def\@tempb {arXiv}\fi \@ifundefined
  {mn@eprint@\@tempb}{\@tempb:\@tempc}{\expandafter \expandafter \csname
  mn@eprint@\@tempb\endcsname \expandafter{\@tempc}}}

\bibitem[\protect\citeauthoryear{{Aladro}, {Mart{\'\i}n},
  {Mart{\'\i}n-Pintado}, {Mauersberger}, {Henkel}, {Oca{\~n}a Flaquer}  \&
  {Amo-Baladr{\'o}n}}{{Aladro} et~al.}{2011}]{aladro2011}
{Aladro} R.,  {Mart{\'\i}n} S.,  {Mart{\'\i}n-Pintado} J.,  {Mauersberger} R.,
  {Henkel} C.,  {Oca{\~n}a Flaquer} B.,   {Amo-Baladr{\'o}n} M.~A.,  2011,
  \mn@doi [\aap] {10.1051/0004-6361/201117397}, \href
  {https://ui.adsabs.harvard.edu/abs/2011A&A...535A..84A} {535, A84}

\bibitem[\protect\citeauthoryear{{Armijos-Abenda{\~n}o}, {Banda-Barrag{\'a}n},
  {Mart{\'\i}n-Pintado}, {D{\'e}nes}, {Federrath}  \&
  {Requena-Torres}}{{Armijos-Abenda{\~n}o} et~al.}{2020}]{armijos2020}
{Armijos-Abenda{\~n}o} J.,  {Banda-Barrag{\'a}n} W.~E.,  {Mart{\'\i}n-Pintado}
  J.,  {D{\'e}nes} H.,  {Federrath} C.,   {Requena-Torres} M.~A.,  2020,
  \mn@doi [\mnras] {10.1093/mnras/staa3119}, \href
  {https://ui.adsabs.harvard.edu/abs/2020MNRAS.499.4918A} {499, 4918}

\bibitem[\protect\citeauthoryear{{Barnes}, {Longmore}, {Battersby}, {Bally},
  {Kruijssen}, {Henshaw}  \& {Walker}}{{Barnes} et~al.}{2017}]{barnes2017}
{Barnes} A.~T.,  {Longmore} S.~N.,  {Battersby} C.,  {Bally} J.,  {Kruijssen}
  J.~M.~D.,  {Henshaw} J.~D.,   {Walker} D.~L.,  2017, \mn@doi [\mnras]
  {10.1093/mnras/stx941}, \href
  {https://ui.adsabs.harvard.edu/abs/2017MNRAS.469.2263B} {469, 2263}

\bibitem[\protect\citeauthoryear{{Belloche}, {M{\"u}ller}, {Menten}, {Schilke}
  \& {Comito}}{{Belloche} et~al.}{2013}]{belloche2013}
{Belloche} A.,  {M{\"u}ller} H.~S.~P.,  {Menten} K.~M.,  {Schilke} P.,
  {Comito} C.,  2013, \mn@doi [\aap] {10.1051/0004-6361/201321096}, \href
  {https://ui.adsabs.harvard.edu/abs/2013A&A...559A..47B} {559, A47}

\bibitem[\protect\citeauthoryear{{Belloche}, {Garrod}, {M{\"u}ller}, {Menten},
  {Medvedev}, {Thomas}  \& {Kisiel}}{{Belloche} et~al.}{2019}]{belloche2019}
{Belloche} A.,  {Garrod} R.~T.,  {M{\"u}ller} H.~S.~P.,  {Menten} K.~M.,
  {Medvedev} I.,  {Thomas} J.,   {Kisiel} Z.,  2019, \mn@doi [\aap]
  {10.1051/0004-6361/201935428}, \href
  {https://ui.adsabs.harvard.edu/abs/2019A&A...628A..10B} {628, A10}

\bibitem[\protect\citeauthoryear{{Belloche} et~al.,}{{Belloche}
  et~al.}{2020}]{belloche2020}
{Belloche} A.,  et~al., 2020, \mn@doi [\aap] {10.1051/0004-6361/201937352},
  \href {https://ui.adsabs.harvard.edu/abs/2020A&A...635A.198B} {635, A198}

\bibitem[\protect\citeauthoryear{Bise, Choi  \& Neumark}{Bise
  et~al.}{1999}]{bise1999}
Bise R.~T.,  Choi H.,   Neumark D.~M.,  1999, The Journal of chemical physics,
  111, 4923

\bibitem[\protect\citeauthoryear{Bise, Hoops  \& Neumark}{Bise
  et~al.}{2001}]{bise2001}
Bise R.~T.,  Hoops A.~A.,   Neumark D.~M.,  2001, The Journal of Chemical
  Physics, 114, 9000

\bibitem[\protect\citeauthoryear{{Blitz}, {Seakins}  \& {Smith}}{{Blitz}
  et~al.}{2009}]{blitz2009}
{Blitz} M.~A.,  {Seakins} P.~W.,   {Smith} I. W.~M.,  2009, \mn@doi [Physical
  Chemistry Chemical Physics (Incorporating Faraday Transactions)]
  {10.1039/b917734e}, \href
  {https://ui.adsabs.harvard.edu/abs/2009PCCP...1110824B} {11, 10824}

\bibitem[\protect\citeauthoryear{{Bonfand}, {Belloche}, {Menten}, {Garrod}  \&
  {M{\"u}ller}}{{Bonfand} et~al.}{2017}]{bonfand2017}
{Bonfand} M.,  {Belloche} A.,  {Menten} K.~M.,  {Garrod} R.~T.,   {M{\"u}ller}
  H.~S.~P.,  2017, \mn@doi [\aap] {10.1051/0004-6361/201730648}, \href
  {https://ui.adsabs.harvard.edu/abs/2017A&A...604A..60B} {604, A60}

\bibitem[\protect\citeauthoryear{{Bonfand}, {Belloche}, {Garrod}, {Menten},
  {Willis}, {St{\'e}phan}  \& {M{\"u}ller}}{{Bonfand}
  et~al.}{2019}]{bonfand2019}
{Bonfand} M.,  {Belloche} A.,  {Garrod} R.~T.,  {Menten} K.~M.,  {Willis} E.,
  {St{\'e}phan} G.,   {M{\"u}ller} H.~S.~P.,  2019, \mn@doi [\aap]
  {10.1051/0004-6361/201935523}, \href
  {https://ui.adsabs.harvard.edu/abs/2019A&A...628A..27B} {628, A27}

\bibitem[\protect\citeauthoryear{{Brack}}{{Brack}}{1999}]{brack1999}
{Brack} A.,  1999, \mn@doi [Advances in Space Research]
  {10.1016/S0273-1177(99)00457-3}, \href
  {https://ui.adsabs.harvard.edu/abs/1999AdSpR..24..417B} {24, 417}

\bibitem[\protect\citeauthoryear{{Chakrabarti} \& {Chakrabarti}}{{Chakrabarti}
  \& {Chakrabarti}}{2000}]{chakrabarti2000}
{Chakrabarti} S.,  {Chakrabarti} S.~K.,  2000, \aap, \href
  {https://ui.adsabs.harvard.edu/abs/2000A&A...354L...6C} {354, L6}

\bibitem[\protect\citeauthoryear{{Chakrabarti}, {Majumdar}, {Das}  \&
  {Chakrabarti}}{{Chakrabarti} et~al.}{2015}]{chakrabarti2015}
{Chakrabarti} S.~K.,  {Majumdar} L.,  {Das} A.,   {Chakrabarti} S.,  2015,
  \mn@doi [\apss] {10.1007/s10509-015-2239-1}, \href
  {https://ui.adsabs.harvard.edu/abs/2015Ap&SS.357...90C} {357, 90}

\bibitem[\protect\citeauthoryear{{Coutens} et~al.,}{{Coutens}
  et~al.}{2018}]{coutens2018}
{Coutens} A.,  et~al., 2018, \mn@doi [\aap] {10.1051/0004-6361/201732346},
  \href {https://ui.adsabs.harvard.edu/abs/2018A&A...612A.107C} {612, A107}

\bibitem[\protect\citeauthoryear{{Cummins}, {Linke}  \& {Thaddeus}}{{Cummins}
  et~al.}{1986}]{cummins1986}
{Cummins} S.~E.,  {Linke} R.~A.,   {Thaddeus} P.,  1986, \mn@doi [\apjs]
  {10.1086/191102}, \href
  {https://ui.adsabs.harvard.edu/abs/1986ApJS...60..819C} {60, 819}

\bibitem[\protect\citeauthoryear{{De Becker}}{{De Becker}}{2013}]{becker2013}
{De Becker} M.,  2013, Bulletin de la Societe Royale des Sciences de Liege,
  \href {https://ui.adsabs.harvard.edu/abs/2013BSRSL..82...33D} {82, 33}

\bibitem[\protect\citeauthoryear{{Dunning}}{{Dunning}}{1989}]{dunning1989}
{Dunning} Thom~H. J.,  1989, \mn@doi [\jcp] {10.1063/1.456153}, \href
  {https://ui.adsabs.harvard.edu/abs/1989JChPh..90.1007D} {90, 1007}

\bibitem[\protect\citeauthoryear{Duvernay, Chiavassa, Borget  \&
  Aycard}{Duvernay et~al.}{2004}]{duvernay2004}
Duvernay F.,  Chiavassa T.,  Borget F.,   Aycard J.-P.,  2004, Journal of the
  American Chemical Society, 126, 7772

\bibitem[\protect\citeauthoryear{Duvernay, Chiavassa, Borget  \&
  Aycard}{Duvernay et~al.}{2005}]{duvernay2005}
Duvernay F.,  Chiavassa T.,  Borget F.,   Aycard J.-P.,  2005, The Journal of
  Physical Chemistry A, 109, 603

\bibitem[\protect\citeauthoryear{{Ellingsen}, {Chen}, {Breen}  \&
  {Qiao}}{{Ellingsen} et~al.}{2017}]{ellingsen2017}
{Ellingsen} S.~P.,  {Chen} X.,  {Breen} S.~L.,   {Qiao} H.~H.,  2017, \mn@doi
  [\mnras] {10.1093/mnras/stx2076}, \href
  {https://ui.adsabs.harvard.edu/abs/2017MNRAS.472..604E} {472, 604}

\bibitem[\protect\citeauthoryear{{Enrique-Romero}, {Rimola}, {Ceccarelli},
  {Ugliengo}, {Balucani}  \& {Skouteris}}{{Enrique-Romero}
  et~al.}{2019}]{enrique2019}
{Enrique-Romero} J.,  {Rimola} A.,  {Ceccarelli} C.,  {Ugliengo} P.,
  {Balucani} N.,   {Skouteris} D.,  2019, \mn@doi [ACS Earth and Space
  Chemistry] {10.1021/acsearthspacechem.9b00156}, \href
  {https://ui.adsabs.harvard.edu/abs/2019ESC.....3.2158E} {3, 2158}

\bibitem[\protect\citeauthoryear{{Espinosa-Garcia}, {Corchado}  \&
  {Sana}}{{Espinosa-Garcia} et~al.}{1993}]{espinosa1993}
{Espinosa-Garcia} J.,  {Corchado} J.,   {Sana} M.,  1993, \mn@doi [Journal de
  Chimie Physique] {10.1051/jcp/1993901181}, \href
  {https://ui.adsabs.harvard.edu/abs/1993JCP....90.1181E} {90, 1181}

\bibitem[\protect\citeauthoryear{{Flower} \& {Pineau des For{\^e}ts}}{{Flower}
  \& {Pineau des For{\^e}ts}}{2015}]{flower2015}
{Flower} D.~R.,  {Pineau des For{\^e}ts} G.,  2015, \mn@doi [\aap]
  {10.1051/0004-6361/201525740}, \href
  {https://ui.adsabs.harvard.edu/abs/2015A&A...578A..63F} {578, A63}

\bibitem[\protect\citeauthoryear{Frisch et~al.,}{Frisch
  et~al.}{2016}]{Gaussian16}
Frisch M.~J.,  et~al., 2016, {Gaussian} 16, Revision A.03.
Gaussian, Inc., Wallingford, CT, USA

\bibitem[\protect\citeauthoryear{Fukui}{Fukui}{1981}]{fukui1981}
Fukui K.,  1981, Accounts of chemical research, 14, 363

\bibitem[\protect\citeauthoryear{{Garrod} \& {Herbst}}{{Garrod} \&
  {Herbst}}{2006}]{garrod2006}
{Garrod} R.~T.,  {Herbst} E.,  2006, \mn@doi [\aap]
  {10.1051/0004-6361:20065560}, \href
  {https://ui.adsabs.harvard.edu/abs/2006A&A...457..927G} {457, 927}

\bibitem[\protect\citeauthoryear{{Garrod}, {Wakelam}  \& {Herbst}}{{Garrod}
  et~al.}{2007}]{garrod2007}
{Garrod} R.~T.,  {Wakelam} V.,   {Herbst} E.,  2007, \mn@doi [\aap]
  {10.1051/0004-6361:20066704}, \href
  {https://ui.adsabs.harvard.edu/abs/2007A&A...467.1103G} {467, 1103}

\bibitem[\protect\citeauthoryear{{Garrod}, {Widicus Weaver}  \&
  {Herbst}}{{Garrod} et~al.}{2008}]{garrod2008}
{Garrod} R.~T.,  {Widicus Weaver} S.~L.,   {Herbst} E.,  2008, \mn@doi [\apj]
  {10.1086/588035}, \href
  {https://ui.adsabs.harvard.edu/abs/2008ApJ...682..283G} {682, 283}

\bibitem[\protect\citeauthoryear{Goerigk \& Grimme}{Goerigk \&
  Grimme}{2011}]{goerigk2011}
Goerigk L.,  Grimme S.,  2011, Journal of chemical theory and computation, 7,
  291

\bibitem[\protect\citeauthoryear{{Graedel}, {Langer}  \& {Frerking}}{{Graedel}
  et~al.}{1982}]{graedel1982}
{Graedel} T.~E.,  {Langer} W.~D.,   {Frerking} M.~A.,  1982, \mn@doi [\apjs]
  {10.1086/190780}, \href
  {https://ui.adsabs.harvard.edu/abs/1982ApJS...48..321G} {48, 321}

\bibitem[\protect\citeauthoryear{{Grimme}}{{Grimme}}{2006}]{grimme2006}
{Grimme} S.,  2006, \mn@doi [\jcp] {10.1063/1.2148954}, \href
  {https://ui.adsabs.harvard.edu/abs/2006JChPh.124c4108G} {124, 034108}

\bibitem[\protect\citeauthoryear{Grimme, Ehrlich  \& Goerigk}{Grimme
  et~al.}{2011}]{grimme2011}
Grimme S.,  Ehrlich S.,   Goerigk L.,  2011, Journal of computational
  chemistry, 32, 1456

\bibitem[\protect\citeauthoryear{{Harada}, {Herbst}  \& {Wakelam}}{{Harada}
  et~al.}{2010}]{harada2010}
{Harada} N.,  {Herbst} E.,   {Wakelam} V.,  2010, \mn@doi [\apj]
  {10.1088/0004-637X/721/2/1570}, \href
  {https://ui.adsabs.harvard.edu/abs/2010ApJ...721.1570H} {721, 1570}

\bibitem[\protect\citeauthoryear{{Hasegawa}, {Herbst}  \& {Leung}}{{Hasegawa}
  et~al.}{1992}]{hasegawa1992}
{Hasegawa} T.~I.,  {Herbst} E.,   {Leung} C.~M.,  1992, \mn@doi [\apjs]
  {10.1086/191713}, \href
  {https://ui.adsabs.harvard.edu/abs/1992ApJS...82..167H} {82, 167}

\bibitem[\protect\citeauthoryear{He, Liu, Lin  \& Melius}{He
  et~al.}{1991}]{He1991}
He Y.,  Liu X.,  Lin M.~C.,   Melius C.~F.,  1991, International Journal of
  Chemical Kinetics, 23, 1129

\bibitem[\protect\citeauthoryear{Jabs, Winnewisser, Belov, Lewen, Maiwald  \&
  Winnewisser}{Jabs et~al.}{1999}]{jabs1999}
Jabs W.,  Winnewisser M.,  Belov S.~P.,  Lewen F.,  Maiwald F.,   Winnewisser
  G.,  1999, Molecular Physics, 97, 213

\bibitem[\protect\citeauthoryear{{J{\o}rgensen} et~al.,}{{J{\o}rgensen}
  et~al.}{2016}]{jorgensen2016}
{J{\o}rgensen} J.~K.,  et~al., 2016, \mn@doi [\aap]
  {10.1051/0004-6361/201628648}, \href
  {https://ui.adsabs.harvard.edu/abs/2016A&A...595A.117J} {595, A117}

\bibitem[\protect\citeauthoryear{{Kawaguchi} et~al.,}{{Kawaguchi}
  et~al.}{1992}]{kawaguchi1992}
{Kawaguchi} K.,  et~al., 1992, \mn@doi [\apjl] {10.1086/186514}, \href
  {https://ui.adsabs.harvard.edu/abs/1992ApJ...396L..49K} {396, L49}

\bibitem[\protect\citeauthoryear{Kilpatrick}{Kilpatrick}{1947}]{kilpatrick1947}
Kilpatrick M.~L.,  1947, Journal of the American Chemical Society, 69, 40

\bibitem[\protect\citeauthoryear{{Ligterink}, {El-Abd}, {Brogan}, {Hunter},
  {Remijan}, {Garrod}  \& {McGuire}}{{Ligterink} et~al.}{2020}]{ligterink2020}
{Ligterink} N. F.~W.,  {El-Abd} S.~J.,  {Brogan} C.~L.,  {Hunter} T.~R.,
  {Remijan} A.~J.,  {Garrod} R.~T.,   {McGuire} B.~M.,  2020, \mn@doi [\apj]
  {10.3847/1538-4357/abad38}, \href
  {https://ui.adsabs.harvard.edu/abs/2020ApJ...901...37L} {901, 37}

\bibitem[\protect\citeauthoryear{Lopez-Sepulcre, Balucani, Ceccarelli, Codella,
  Dulieu  \& Theule}{Lopez-Sepulcre et~al.}{2019}]{lopez2019}
Lopez-Sepulcre A.,  Balucani N.,  Ceccarelli C.,  Codella C.,  Dulieu F.,
  Theule P.,  2019, ACS Earth and Space Chemistry, 3, 2122

\bibitem[\protect\citeauthoryear{{Marcelino} et~al.,}{{Marcelino}
  et~al.}{2018}]{marcelino2018}
{Marcelino} N.,  et~al., 2018, \mn@doi [\aap] {10.1051/0004-6361/201731955},
  \href {https://ui.adsabs.harvard.edu/abs/2018A&A...620A..80M} {620, A80}

\bibitem[\protect\citeauthoryear{{Mart{\'\i}n}, {Mauersberger},
  {Mart{\'\i}n-Pintado}, {Henkel}  \& {Garc{\'\i}a-Burillo}}{{Mart{\'\i}n}
  et~al.}{2006}]{martin2006}
{Mart{\'\i}n} S.,  {Mauersberger} R.,  {Mart{\'\i}n-Pintado} J.,  {Henkel} C.,
   {Garc{\'\i}a-Burillo} S.,  2006, \mn@doi [\apjs] {10.1086/503297}, \href
  {https://ui.adsabs.harvard.edu/abs/2006ApJS..164..450M} {164, 450}

\bibitem[\protect\citeauthoryear{{McGuire} et~al.,}{{McGuire}
  et~al.}{2012}]{mcguire2012}
{McGuire} B.~A.,  et~al., 2012, \mn@doi [\apjl] {10.1088/2041-8205/758/2/L33},
  \href {https://ui.adsabs.harvard.edu/abs/2012ApJ...758L..33M} {758, L33}

\bibitem[\protect\citeauthoryear{{Merz}, {Aguiar}  \& {da Silva}}{{Merz}
  et~al.}{2014}]{merz2014}
{Merz} Kenneth~M. J.,  {Aguiar} E.~C.,   {da Silva} J. B.~P.,  2014, \mn@doi
  [Journal of Physical Chemistry A] {10.1021/jp5018778}, \href
  {https://ui.adsabs.harvard.edu/abs/2014JPCA..118.3637M} {118, 3637}

\bibitem[\protect\citeauthoryear{{Nummelin}, {Bergman}, {Hjalmarson},
  {Friberg}, {Irvine}, {Millar}, {Ohishi}  \& {Saito}}{{Nummelin}
  et~al.}{2000}]{nummelin2000}
{Nummelin} A.,  {Bergman} P.,  {Hjalmarson} {\r{A}}.,  {Friberg} P.,  {Irvine}
  W.~M.,  {Millar} T.~J.,  {Ohishi} M.,   {Saito} S.,  2000, \mn@doi [\apjs]
  {10.1086/313376}, \href
  {https://ui.adsabs.harvard.edu/abs/2000ApJS..128..213N} {128, 213}

\bibitem[\protect\citeauthoryear{{{\"O}berg}, {Fuchs}, {Awad}, {Fraser},
  {Schlemmer}, {van Dishoeck}  \& {Linnartz}}{{{\"O}berg}
  et~al.}{2007}]{oberg2007}
{{\"O}berg} K.~I.,  {Fuchs} G.~W.,  {Awad} Z.,  {Fraser} H.~J.,  {Schlemmer}
  S.,  {van Dishoeck} E.~F.,   {Linnartz} H.,  2007, \mn@doi [\apjl]
  {10.1086/519281}, \href
  {https://ui.adsabs.harvard.edu/abs/2007ApJ...662L..23O} {662, L23}

\bibitem[\protect\citeauthoryear{{Palau} et~al.,}{{Palau}
  et~al.}{2017}]{palau2017}
{Palau} A.,  et~al., 2017, \mn@doi [\mnras] {10.1093/mnras/stx004}, \href
  {https://ui.adsabs.harvard.edu/abs/2017MNRAS.467.2723P} {467, 2723}

\bibitem[\protect\citeauthoryear{Papajak, Leverentz, Zheng  \& Truhlar}{Papajak
  et~al.}{2009}]{papajak2009}
Papajak E.,  Leverentz H.~R.,  Zheng J.,   Truhlar D.~G.,  2009, Journal of
  chemical theory and computation, 5, 1197

\bibitem[\protect\citeauthoryear{Purvis~III \& Bartlett}{Purvis~III \&
  Bartlett}{1982}]{purvis1982}
Purvis~III G.~D.,  Bartlett R.~J.,  1982, The Journal of Chemical Physics, 76,
  1910

\bibitem[\protect\citeauthoryear{{Puzzarini}, {Salta}, {Tasinato}, {Lupi},
  {Cavallotti}  \& {Barone}}{{Puzzarini} et~al.}{2020}]{puzzarini2020}
{Puzzarini} C.,  {Salta} Z.,  {Tasinato} N.,  {Lupi} J.,  {Cavallotti} C.,
  {Barone} V.,  2020, \mn@doi [\mnras] {10.1093/mnras/staa1652}, \href
  {https://ui.adsabs.harvard.edu/abs/2020MNRAS.496.4298P} {496, 4298}

\bibitem[\protect\citeauthoryear{{Quan} \& {Herbst}}{{Quan} \&
  {Herbst}}{2007}]{quan2007}
{Quan} D.,  {Herbst} E.,  2007, \mn@doi [\aap] {10.1051/0004-6361:20078246},
  \href {https://ui.adsabs.harvard.edu/abs/2007A&A...474..521Q} {474, 521}

\bibitem[\protect\citeauthoryear{{Quan}, {Herbst}, {Osamura}  \&
  {Roueff}}{{Quan} et~al.}{2010}]{quan2010}
{Quan} D.,  {Herbst} E.,  {Osamura} Y.,   {Roueff} E.,  2010, \mn@doi [\apj]
  {10.1088/0004-637X/725/2/2101}, \href
  {https://ui.adsabs.harvard.edu/abs/2010ApJ...725.2101Q} {725, 2101}

\bibitem[\protect\citeauthoryear{{Quan}, {Herbst}, {Corby}, {Durr}  \&
  {Hassel}}{{Quan} et~al.}{2016}]{quan2016}
{Quan} D.,  {Herbst} E.,  {Corby} J.~F.,  {Durr} A.,   {Hassel} G.,  2016,
  \mn@doi [\apj] {10.3847/0004-637X/824/2/129}, \href
  {https://ui.adsabs.harvard.edu/abs/2016ApJ...824..129Q} {824, 129}

\bibitem[\protect\citeauthoryear{{Rathborne} et~al.,}{{Rathborne}
  et~al.}{2015}]{rathborne2015}
{Rathborne} J.~M.,  et~al., 2015, \mn@doi [\apj] {10.1088/0004-637X/802/2/125},
  \href {https://ui.adsabs.harvard.edu/abs/2015ApJ...802..125R} {802, 125}

\bibitem[\protect\citeauthoryear{{Rivilla} et~al.,}{{Rivilla}
  et~al.}{2021}]{rivilla2021}
{Rivilla} V.~M.,  et~al., 2021, \mn@doi [\mnras] {10.1093/mnrasl/slab074},
  \href {https://ui.adsabs.harvard.edu/abs/2021MNRAS.506L..79R} {506, L79}

\bibitem[\protect\citeauthoryear{{Rodr{\'\i}guez-Fern{\'a}ndez},
  {Mart{\'\i}n-Pintado}, {Fuente}  \& {Wilson}}{{Rodr{\'\i}guez-Fern{\'a}ndez}
  et~al.}{2004}]{rodriguez2004}
{Rodr{\'\i}guez-Fern{\'a}ndez} N.~J.,  {Mart{\'\i}n-Pintado} J.,  {Fuente} A.,
   {Wilson} T.~L.,  2004, \mn@doi [\aap] {10.1051/0004-6361:20041370}, \href
  {https://ui.adsabs.harvard.edu/abs/2004A&A...427..217R} {427, 217}

\bibitem[\protect\citeauthoryear{{Ruaud}, {Wakelam}  \& {Hersant}}{{Ruaud}
  et~al.}{2016}]{ruaud2016}
{Ruaud} M.,  {Wakelam} V.,   {Hersant} F.,  2016, \mn@doi [\mnras]
  {10.1093/mnras/stw887}, \href
  {https://ui.adsabs.harvard.edu/abs/2016MNRAS.459.3756R} {459, 3756}

\bibitem[\protect\citeauthoryear{Scuseria \& Schaefer~III}{Scuseria \&
  Schaefer~III}{1989}]{scuseria1989}
Scuseria G.~E.,  Schaefer~III H.~F.,  1989, The Journal of Chemical Physics,
  90, 3700

\bibitem[\protect\citeauthoryear{Scuseria, Janssen  \& Schaefer~Iii}{Scuseria
  et~al.}{1988}]{scuseria1988}
Scuseria G.~E.,  Janssen C.~L.,   Schaefer~Iii H.~F.,  1988, The Journal of
  Chemical Physics, 89, 7382

\bibitem[\protect\citeauthoryear{{Skouteris}, {Vazart}, {Ceccarelli},
  {Balucani}, {Puzzarini}  \& {Barone}}{{Skouteris}
  et~al.}{2017}]{skouteris2017}
{Skouteris} D.,  {Vazart} F.,  {Ceccarelli} C.,  {Balucani} N.,  {Puzzarini}
  C.,   {Barone} V.,  2017, \mn@doi [\mnras] {10.1093/mnrasl/slx012}, \href
  {https://ui.adsabs.harvard.edu/abs/2017MNRAS.468L...1S} {468, L1}

\bibitem[\protect\citeauthoryear{{Sleiman}, {El Dib}, {Talbi}  \&
  {Canosa}}{{Sleiman} et~al.}{2018a}]{sleiman2018a}
{Sleiman} C.,  {El Dib} G.,  {Talbi} D.,   {Canosa} A.,  2018a, \mn@doi [ACS
  Earth and Space Chemistry] {10.1021/acsearthspacechem.8b00098}, \href
  {https://ui.adsabs.harvard.edu/abs/2018ESC.....2.1047S} {2, 1047}

\bibitem[\protect\citeauthoryear{Sleiman, El~Dib, Rosi, Skouteris, Balucani  \&
  Canosa}{Sleiman et~al.}{2018b}]{sleiman2018b}
Sleiman C.,  El~Dib G.,  Rosi M.,  Skouteris D.,  Balucani N.,   Canosa A.,
  2018b, Physical Chemistry Chemical Physics, 20, 5478

\bibitem[\protect\citeauthoryear{{Smith}, {Talbi}  \& {Herbst}}{{Smith}
  et~al.}{2001}]{smith2001}
{Smith} I.~W.~M.,  {Talbi} D.,   {Herbst} E.,  2001, \mn@doi [\aap]
  {10.1051/0004-6361:20010126}, \href
  {https://ui.adsabs.harvard.edu/abs/2001A&A...369..611S} {369, 611}

\bibitem[\protect\citeauthoryear{{Smith}, {Herbst}  \& {Chang}}{{Smith}
  et~al.}{2004}]{smith2004}
{Smith} I. W.~M.,  {Herbst} E.,   {Chang} Q.,  2004, \mn@doi [\mnras]
  {10.1111/j.1365-2966.2004.07656.x}, \href
  {https://ui.adsabs.harvard.edu/abs/2004MNRAS.350..323S} {350, 323}

\bibitem[\protect\citeauthoryear{Steinman, Lemmon  \& Calvin}{Steinman
  et~al.}{1964}]{steinman1964}
Steinman G.,  Lemmon R.~M.,   Calvin M.,  1964, Proceedings of the National
  Academy of Sciences of the United States of America, 52, 27

\bibitem[\protect\citeauthoryear{{Talbi} \& {Smith}}{{Talbi} \&
  {Smith}}{2009}]{talbi2009}
{Talbi} D.,  {Smith} I. W.~M.,  2009, \mn@doi [Physical Chemistry Chemical
  Physics (Incorporating Faraday Transactions)] {10.1039/B908416A}, \href
  {https://ui.adsabs.harvard.edu/abs/2009PCCP...11.8477T} {11, 8477}

\bibitem[\protect\citeauthoryear{Tordini, Bencini, Bruschi, De~Gioia, Zampella
  \& Fantucci}{Tordini et~al.}{2003}]{tordini2003}
Tordini F.,  Bencini A.,  Bruschi M.,  De~Gioia L.,  Zampella G.,   Fantucci
  P.,  2003, The Journal of Physical Chemistry A, 107, 1188

\bibitem[\protect\citeauthoryear{{Turner}, {Liszt}, {Kaifu}  \&
  {Kisliakov}}{{Turner} et~al.}{1975}]{turner1975}
{Turner} B.~E.,  {Liszt} H.~S.,  {Kaifu} N.,   {Kisliakov} A.~G.,  1975,
  \mn@doi [\apjl] {10.1086/181963}, \href
  {https://ui.adsabs.harvard.edu/abs/1975ApJ...201L.149T} {201, L149}

\bibitem[\protect\citeauthoryear{{White}, {Araki}, {Greaves}, {Ohishi}  \&
  {Higginbottom}}{{White} et~al.}{2003}]{white2003}
{White} G.~J.,  {Araki} M.,  {Greaves} J.~S.,  {Ohishi} M.,   {Higginbottom}
  N.~S.,  2003, \mn@doi [\aap] {10.1051/0004-6361:20030841}, \href
  {https://ui.adsabs.harvard.edu/abs/2003A&A...407..589W} {407, 589}

\bibitem[\protect\citeauthoryear{Williams \& Ibrahim}{Williams \&
  Ibrahim}{1981}]{williams1981}
Williams A.,  Ibrahim I.~T.,  1981, Chemical Reviews, 81, 589

\bibitem[\protect\citeauthoryear{Woon \& Herbst}{Woon \&
  Herbst}{2009}]{woon2009}
Woon D.~E.,  Herbst E.,  2009, \mn@doi [Astrophysical Journal Supplement
  Series] {10.1088/0067-0049/185/2/273}, 185, 273

\bibitem[\protect\citeauthoryear{Yadav, Misra, Tandon  et~al.}{Yadav
  et~al.}{2019}]{yadav2019}
Yadav M.,  Misra A.,  Tandon P.,   et~al., 2019, Origins of Life and Evolution
  of Biospheres, 49, 89

\bibitem[\protect\citeauthoryear{{Zeng} et~al.,}{{Zeng}
  et~al.}{2018}]{zeng2018}
{Zeng} S.,  et~al., 2018, \mn@doi [\mnras] {10.1093/mnras/sty1174}, \href
  {https://ui.adsabs.harvard.edu/abs/2018MNRAS.478.2962Z} {478, 2962}

\bibitem[\protect\citeauthoryear{{Zeng}, {Qu{\'e}nard}, {Jim{\'e}nez-Serra},
  {Mart{\'\i}n-Pintado}, {Rivilla}, {Testi}  \&
  {Mart{\'\i}n-Dom{\'e}nech}}{{Zeng} et~al.}{2019}]{zeng2019}
{Zeng} S.,  {Qu{\'e}nard} D.,  {Jim{\'e}nez-Serra} I.,  {Mart{\'\i}n-Pintado}
  J.,  {Rivilla} V.~M.,  {Testi} L.,   {Mart{\'\i}n-Dom{\'e}nech} R.,  2019,
  \mn@doi [\mnras] {10.1093/mnrasl/slz002}, \href
  {https://ui.adsabs.harvard.edu/abs/2019MNRAS.484L..43Z} {484, L43}

\bibitem[\protect\citeauthoryear{{Zeng} et~al.,}{{Zeng}
  et~al.}{2020}]{zeng2020}
{Zeng} S.,  et~al., 2020, \mn@doi [\mnras] {10.1093/mnras/staa2187}, \href
  {https://ui.adsabs.harvard.edu/abs/2020MNRAS.497.4896Z} {497, 4896}

\bibitem[\protect\citeauthoryear{{Zhang}, {Quan}, {Chang}, {Herbst}, {Esimbek}
  \& {Webb}}{{Zhang} et~al.}{2020}]{zhang2020}
{Zhang} X.,  {Quan} D.,  {Chang} Q.,  {Herbst} E.,  {Esimbek} J.,   {Webb} M.,
  2020, \mn@doi [\mnras] {10.1093/mnras/staa1979}, \href
  {https://ui.adsabs.harvard.edu/abs/2020MNRAS.497..609Z} {497, 609}

\makeatother
\end{thebibliography}








\bsp	
\label{lastpage}
\end{document}